\documentstyle[11pt,psfig]{article}

\newlength{\figwidth}
\newlength{\figwidthb}
\newcommand{\be}{\begin{equation}}
\newcommand{\ee}{\end{equation}}
\newcommand{\bea}{\begin{eqnarray}}
\newcommand{\eea}{\end{eqnarray}}
\newcommand{\no}{\noindent}

\newcommand{\sss}{\scriptscriptstyle}

\newcommand{\NPB}[3]{Nucl. Phys. B {\bf #1} (#2) #3}
\newcommand{\PRD}[3]{Phys. Rev. D {\bf #1} (#2) #3}
\newcommand{\PRL}[3]{Phys. Rev. Lett. {\bf #1} (#2) #3}

\newcommand{\NPBPS}[3]{Nucl. Phys. B (Proc. Suppl.) {\bf #1} (#2) #3}

\newcommand{\PR}[3]{Phys. Rep. {\bf #1} (#2) #3}

\newcommand{\towh}[3]{
{\!\begin{array}{c}{#1}\vspace*{-0.2cm}\\{#2}\vspace*{-0.2cm}\\{#3}\end{array}\!}}

\title{Meson correlators in finite temperature lattice QCD}

\author{{\sl QCD-TARO Collaboration:} \smallskip \\
 Ph.~de~Forcrand$^1$, 
 M.~Garc{\'\i}a~P\'erez$^2$, 
 T.~Hashimoto$^3$, 
 S.~Hioki$^4$, 
 H.~Matsufuru$^5$, \\
 O.~Miyamura$^6$, 
 A.~Nakamura$^7$,
 I.-O.~Stamatescu$^{8,9}$, 
 T. Takaishi$^{10}$ 
 and T.~Umeda$^6$   \bigskip \\
{\small $^1$ETH-Z\"urich, CH-8092 Z\"urich, Switzerland}\\
{\small $^2$Dept. F\'{\i}sica Te\'orica, Universidad Aut\'onoma de Madrid,
              E-28049 Madrid, Spain}\\
{\small $^3$Department of Applied Physics, Faculty of Engineering,}\\
            {\small Fukui University, Fukui 910-8507, Japan}\\
{\small $^4$Department of Physics, Tezukayama University,
                Nara 631-8501, Japan}\\
{\small $^5$Research Center for Nuclear Physics, Osaka University,
            Ibaraki 567-0047, Japan}\\
{\small $^6$Department of Physics, Hiroshima University,
                  Higashi-Hiroshima 739-8526, Japan}\\
{\small $^7$Research Institute for Information Science and Education,
                   Hiroshima University,}\\
            {\small Higashi-Hiroshima 739-8521, Japan}\\
{\small $^8$Institut f\"ur Theoretische Physik, Universit\"at Heidelberg,
                    D-69120 Heidelberg, Germany}\\
{\small $^9$FEST, Schmeilweg 5, D-69118 Heidelberg, Germany}\\
{\small $^{10}$Hiroshima University of Economics, Hiroshima 731-01, Japan}
}

\date{\today \bigskip}

\begin{document}

\maketitle

\begin{abstract} We analyze {\it temporal}  and
{\it spatial} meson correlators in quenched lattice QCD
at $T \geq 0$. Below $T_c$ we observe little change in the meson
properties as compared with $T=0$. Above $T_c$ we observe new
features: chiral symmetry restoration and signals of
plasma formation, but also indication of persisting
``mesonic" (metastable) states and different temporal and  
spatial  ``masses"
in the mesonic channels. This suggests a complex picture of QGP in the
region  $1\, - \, 1.5\, T_c$.
\end{abstract}


\clearpage

\section{Introduction}

With increasing temperature, hadronic correlators are expected
to change their nature drastically (see, e.g., \cite{hild}, \cite{FTreviews}).
At the critical temperature, the deconfinement of color degrees
of freedom and the restoration of the chiral symmetry are
expected to occur simultaneously.
Two ``extreme'' pictures are frequently used to describe the
low and the high $T$ regimes, respectively:
the weakly interacting meson gas, where we expect the
mesons to become effective resonance modes with a small mass shift
and width due to the interaction;
and the perturbative quark gluon plasma (QGP), where the mesons should eventually
disappear and (at very high $T$) perturbative effects should dominate.

Near to the critical temperature, however, the actual physical situation
is more involved. The interaction with a hot meson gas and with baryonic matter
have been studied in various phenomenological models which predict appreciable
changes in the vector meson properties (see, e.\,g., \cite{rwb}).
In a NJL model \cite{HK94} the scalar and the
pseudo-scalar modes are found to 
correlate strongly, and to subsist even above the transition
as so-called ``soft modes'', corresponding to
narrow peaks in the spectral function and realized as the fluctuation
of the order parameter of the chiral symmetry restoration transition.
On the other hand, at the short distance scale, the fundamental
excitation should be quarks and gluons.
Lattice QCD results on the quark number susceptibility support this
view \cite{Got87}. These pictures may not be contradicting each other:
DeTar conjectured the existence of
excitations in the QGP phase corresponding to
different distance scales, and pointed out to the possibility of
``confinement'' still ruling the large distance scales
\cite{dTar}. While with increasing temperature the physics should
appear increasingly dominated by quark and gluon degrees of freedom,
in accordance to the perturbative high temperature picture, 
the intermediate temperature range  above $T_c$ seems to be
much more complex and dominated by strongly interacting quarks 
(see also \cite{biel1}), which, as we shall see below, even tend
to stay strongly spatially correlated and thus agree with a picture of
effective, low energy modes in the mesonic channels. 
Since these matters are related to questions about the evolution of the early
universe, on one hand, and to the interesting 
results from heavy ion collision experiments \cite{hi},
where QGP conditions  are
being realized \cite{hipr},
on the other hand, it is important to have quantitative estimates
 in addition to qualitative
understanding and we
need  model-independent studies of the hadronic
correlators at finite temperature.

Lattice simulations are the most powerful instrument at present 
 to investigate
such problems in the fundamental theoretical framework of quantum
chromodynamics (QCD). Extensive studies have been
dedicated to the thermodynamics of the finite temperature
transition (see, e.\,g.,  \cite{LattFT},\cite{LattFTl}). Concerning the hadronic
sectors, numerical analysis of ``screening" (spatial) propagators
indicate correlated (bound?) quarks while the
mesonic ``screening masses" increase toward the 2-free quark
threshold ($2\pi T$, induced by the anti-periodic boundary conditions in
the temporal direction) \cite{scr}. Since the spatial directions 
can (and must) be 
made large,
these propagators are unproblematic in principle 
and can be studied as well as
for $T=0$: at any $T>0$ the propagation in the space directions
represents  in fact a   $T=0$ problem 
(with asymmetric finite size effects).
The interpretation of the results from
``screening" correlators in terms of modes of the  temporal dynamics
is far from straightforward, however:
since in the Euclidean formulation the $O(4)$ symmetry is
broken at $T>0$ (see, e.g., \cite{LvW87}),
 physics appears different depending on whether we probe the space 
(``$\sigma$": $\vec{x}$) or time  (``$\tau$": $t$)
 direction. The static quark potential associated to propagation in
a spatial direction, for example, is a very anisotropic quantity
which above $T_c$ still grows linearly in two of the spatial directions
(confines), in contrast to the potential associated to
the propagation in the $t-$direction which is isotropic and not confining.
Therefore we need to investigate hadronic
correlators with full ``space-time" structure, in particular the
propagation in the Euclidean time.

Ideally, what we should like to do is to reconstruct the spectral 
function in the given channel. Then we could directly compare with the results
from heavy-ions experiments, see e.\,g., \cite{hi}.
  The spectral function at finite temperature can be extracted from the
correlators in the (Euclidean) temporal direction whose
 extent $l_{\tau}$ is related to the temperature as $T=1/l_{\tau}$
\cite{LvW87, Abr59}.
These data (after Fourier transforming the 
 $t$-correlators) are given at  discrete Matsubara frequencies 
on the imaginary energy axis and are affected by errors. The
extraction of the spectral function implies (logically)
an interpolation and an analytic continuation to the real energy axis.
For a numerical analysis, which produces a finite amount of information,
this is an
improperly posed problem. Its solution is dependent on imposing
supplementary conditions (``a priori information") to regularize
the algorithms and to prevent the amplification of the errors.
These conditions can be either based on general, statistical arguments 
(e.g. variance limitation, Bayesian analysis, maximal entropy method)
or on particular, physical expectations (e.g., using an ansatz for the
spectral  function which leads to an explicit analytic form for
the correlator, to be fitted against the data).
We should, however, be aware of the fact that all regularization introduces
a bias and therefore this problem is fundamentally intricate.

The main difficulty in the numerical calculation at $T>0$
originates in the short
temporal extent $ l_{\tau}= 1/T$. Beyond the general necessity of
 producing enough and precise data the 
$T>0$ problem is doubly complicated as compared to the $T=0$ one:
on one hand the structure we may expect is more complex than just a pole,
on the other hand the time extent of the propagation cannot be made large 
to select the low energy contributions. We shall now briefly discuss 
 this questions and thereby also introduce our procedure to deal with them.

1) {\it Lattice problems:} Large $T$ can be
achieved  using small
$N_{\tau} = l_{\tau}/a$ ($a$: lattice spacing), 
however this leads to systematic errors \cite{fthl}.
Moreover, having the
$t$-propagators at only a few ($N_{\tau}/2$) points makes it difficult
to characterize the unknown structure in the corresponding channels:
practically any ansatz can be fitted through 2-3 points!
To  obtain a fine
$t$-discretization
and thus  detailed
 $t$-correlators, while avoiding
prohibitively large lattices (we need large spatial size in order to
avoid finite size effects, typically $l_{\sigma} \sim 3 l_{\tau}$),
we proposed \cite{HNS93a, HNS93b} to use 
different lattice spacings
in space  and in time,
$a_{\sigma}/a_{\tau}=\xi >1 $ \cite{Kar82}.
The renormalization analysis of such lattices, however, is more involved,
because of the supplementary parameter $\xi$ and this introduces also
some uncertainties.

2) {\it Physical problems:} The 
low energy structure of the mesonic channels
cannot be observed directly,
due to the inherently coarse resolution $2\pi T = 2\pi /l_{\tau}$ 
of the imaginary {\it energy}  axis. Refining the discretization of the 
{\it time} axis improves the fitting and analytic continuation problem, but
although we are following  the question of the spectral analysis
we do not have yet reliable results at $T>0$ for this challenging question.
Our problem setting here is therefore more limited: we shall try to
recognize mesonic states and ask about their character and properties
at various temperatures.

In this paper, we investigate the full four-dimensional structure of
the meson correlators on an anisotropic lattice in the quenched
approximation.
Thus the phase transition is the deconfining one, and the hadronic
correlators are constructed with quark propagators on the background
gauge field. 

Our strategy here is  
the following: we first select the mesonic ground states of the $T=0$
problem (where the time direction can be made sufficiently large --
about 3.2 fm in our case, which, at the quark
masses we work with, means about 8 pion correlation lengths) 
and characterize their internal structure by
measuring the (Coulomb gauge) wave functions. Then we ask whether states
characterized by similar internal structure can be retrieved at higher
temperature, try to reconstruct 
them with help of correspondingly smeared sources and investigate 
how they are affected by the temperature.
If
 the changes in the correlators
are small, which is consistent with mesons interacting weakly with other
hadron-like modes in the thermal bath, this procedure allows us
to define ``effective modes". 
Large changes  will signal the
breakdown of this weakly interacting gas picture  and there
we must  try to compare our observations
with other pictures, in particular the perturbative QGP.

The meson correlators in the temporal direction play a central 
role in this study, which is therefore meant to supplement other approaches, 
including studies of
screening propagators \cite{scr}, \cite{scrch}. To understand the effect of fixing a mesonic source 
we employ three kind of meson operator smearing.
The propagators and the wave functions are also 
compared with those of 
mesons composed of free quark propagators (``free" mesons) \footnote{These
can be seen
as  quark-antiquark correlation functions in the corresponding
meson channels in lowest order perturbation theory.}. 
Finally we attempt a chiral limit;
note, however,  that even with anisotropic lattices the short 
physical extent in the temporal direction makes the quantitative
estimate of the (temporal) masses (if they exist) difficult.
Our program should not be understood as an alternative for a study
of the spectral function at $T > 0$, 
but as an attempt to answer some special questions about
the phenomena in QGP. In that sense our results only offer partial views.

To prevent a certain confusion we stress here that we do not look
for the eigenstates of the Hamiltonian (transfer matrix), which show up
as asymptotic states for $t \rightarrow \infty$ at  $T=0$. 
At non-zero temperature the  
 physical processes are
essentially dependent on the mixtures induced by the thermal bath.
For the
physical picture and for building models the question is whether these
phenomena can be described in terms of some effective excitations 
(quasi-particles \cite{thir}), which ``replace" thus the fundamental
 particle modes,
or completely new states dominate the physics above some $T$.

Our analysis proceeds in three steps: 
\begin{itemize}
\item[1.] Analysis of temporal propagators. Here we  try to
see what kind of excitations propagate in the mesonic channels at $T>0$
based on the $t-$dependence of these propagators (``effective mass" $m(t)$).
\item[2.]  Analysis of the ``Coulomb gauge wave functions". Here we
study the behavior of the  temporal correlators with the distance between
$q$ and ${\bar q}$ at the sink, which
provides us with information of the spatial correlation between the quarks
at given $t$.
\item[3.]  Analysis of the temperature dependence of the temporal and spatial
masses of the putative states which are compatible with the
behavior observed at the previous steps. 
\end{itemize}

Note that due to the quenched approximation the dynamics is incomplete.
In a strong sense the Hamiltonian does not posses true mesonic states
and only provides the gluonic interactions responsible for the forces
binding the quarks. This is not a problem specific to nonzero temperature 
but 
is the same already at $T=0$. The success of the spectrum calculations 
at $T=0$ indicates, however, that one should not consider 
quenched QCD as a theory by itself but as an approximation to the full
theory (which posses genuine asymptotic mesonic states) and the exclusion
of $q$-$\bar{q}$ pair creation as a reasonably small error at least concerning
some of the characteristics of the hadrons.
In particular,  we observe strong  indication of  chiral symmetry
restoration above the transition temperature.

The paper is organized as follows.
In the next section we describe our analysis 
strategy in some more detail.
Section~\ref{sec:Setup} describes the preparation of the lattice:
the introduction of anisotropic lattice actions and the simulation
parameters (the ``calibration'', i.e. the tuning of the anisotropy parameters,
is described in detail in the Appendix).
The subsequent two sections present the results of the simulation:
In Section~4 we observe the correlators at zero temperature
and discuss the source smearing and the variational analysis.
The results at finite temperature are presented in Section~\ref{sec:FT}.
The last section is reserved for discussion and outlook.

\section{Analysis strategy}

\subsection{Comments on the physical problems}

We here should like to illustrate the problems raised by
the finite temperature  
and the question of the source in the frame of our approach. The reader
who is familiar with these problems may skip this section. 

Let us consider that we use some meson operator $\Phi$,
then the
propagator at $\beta \equiv 1/T < \infty$ in Euclidean time $t>0$ is:
\bea
G^{(\Phi\rightarrow \Phi)}_{\beta}(t) &=& \langle \Phi(t)\Phi(0)
\rangle =  \frac{1}{Z} \int_{(a)pbc}
\Phi(t)\Phi(0)
{\rm e}^{-\int_0^{\beta}{\cal L}} \\
&=&  \frac{1}{Z} {\rm Tr}\left[
{\cal T}^{N_{\tau}-t} \Phi {\cal T}^t \Phi \right]    
=  {\rm Tr}\left[ {\rm e}^{-\beta\, H}{\rm e}^{ t\, H} 
\Phi{\rm e}^{-t\, H}\Phi\right] /
{\rm Tr}\left[ {\rm e}^{-\beta\, H}\right]  \label{e.tgf1} \\
 &=& \sum_{n,k}{\rm e}^{-(\beta-t)\, E_n - t\,E_k}
\langle \phi_n |\Phi | \phi_k \rangle
\langle \phi_k |\Phi |\phi_n \rangle / \sum_n{\rm e}^{-\beta\, E_n}
\label{e.tgf2}
\\
&=& \sum_{n,k} {\rm e}^{-\frac{\beta}{2} (E_n+E_k)} c_{nk}^2
{\rm cosh}\left[ (\beta/2-t)(E_k-E_n)\right ]/ \sum_n{\rm e}^{-\beta\, E_n}
\label{e.tgf3}
\eea
where ${\phi_n}$ are eigenstates of the Hamiltonian representing, say,
(multi-)meson states and other hadron-like modes and we have:
\be
\langle \phi_n |\Phi | \phi_k \rangle = c_{nk},\ \  c_{kn}=c^{*}_{nk}=c_{nk}
\label{e.tgf4}
\ee
Here we expressed 
everything in units of $a_{\tau} (a_{\tau}^{-1})$, hence  
$\beta=N_{\tau}$; ${\cal T} = {\rm exp}(-H)$ is the transfer matrix. 
For $T=0$ only the vacuum
$n=0$ survives in the  sum over $n$ in (\ref{e.tgf2}). Assume each $\Phi$  selects 
not only the mesonic ground state, say $| \phi_1 \rangle$,
 but also some other, excited 
states, then 
\be
\langle \phi_k |\Phi | \phi_0 \rangle \simeq c_{01} \delta_{k,1} +
c_{02} \delta_{k,2} + \dots
\label{e.ft0}
\ee
and the zero temperature propagator is (we put $E_0=0$ for simplicity): 
\be
G^{(\Phi\rightarrow \Phi)}_{\infty}(t) \simeq c_{01}^2 {\rm e}^{-t\, E_1} 
+ c_{02}^2  {\rm e}^{-t\, E_2} + \dots
\label{e.pt0}
\ee
Hence the lightest state contribution will dominate at large $t$.
Tuning a ``perfect" source at $T=0$ we ideally achieve $c_{0k}=0$ for $k \ne 1$
and thus see only this contribution at all $t$. 
Suppose that we have been able to
construct in this way a ``perfect" operator $\Phi_1 \sim a+a^{\dag}$, with
$a(a^{\dag})$  the annihilation (creation) operator for
a meson in the ground state. 
Then at $T=0$ $G$ reduces to the first term,
as desired

\be
G^{(\Phi_1\rightarrow \Phi_1)}_{\infty}(t) \simeq  c_{01}^2{\rm e}^{-t\, E_1}
\label{e.ts0}
\ee 
(note that correlators with different operators at the source and the sink
also project only on the ground state if either the source or the sink is 
``perfect").
With increasing $T$, however, further states
beyond the vacuum  survive  
in the sum over $n$ and  acting on each of them $\Phi_1$ ``adds/subtracts" 
a meson
to whatever is there, correspondingly selecting  from the inner sum the
states $k$ onto which this new state projects, 
$\langle \phi_n |\Phi_1 | \phi_k \rangle \ne 0$, 
in a sloppy notation $k \in
\{n_{\pm1}\}$.
Instead of (\ref{e.ts0}) the correlator is now  
a sum of contributions and  
we ask whether this mixture
can be described by an effective mode $|{\tilde {\phi}}_1\rangle$ of
energy ${\tilde {E}}_1^{({\beta})}$ such that we can write,
similarly to the $T=0$ expression:
\bea
G^{(\Phi_1\rightarrow \Phi_1)}_{\beta}(t) 
&=&
\sum_n \sum_{k \in \{n_{\pm 1}\}}{\rm e}^{-\frac{\beta}{2} (E_n+E_k)}
c_{nk}^2 \cosh \left[\left (\frac{\beta}{2}-t\right)(E_k-E_n)\right ]
/\sum_n{\rm e}^{-\beta\, E_n} \nonumber \\
&\sim& {\rm cosh}\left[\left ({{\beta}\over 2} - t \right)
{\tilde E}_{1}^{({\beta})}\right] .
\label{e.effm}
\eea

To fix the ideas let us consider an 
oscillator with frequency $\omega$ and a small anharmonic perturbation:
\be
H= H_0 + \epsilon H_{\rm I} 
\label{e.hosc}
\ee
(this may be considered a caricature of a weakly interacting 
meson gas, say). To first order in $\epsilon$ we can use the
unperturbed basis to calculate the propagator $G(t)$.
Let $\Phi_1$ be the ground state operator,
\be
  \Phi_1 = (a + a^{\dag})/\sqrt{2\omega} \ , \ \
\langle \phi_k |\Phi_1^{\dag}|\phi_n \rangle = \frac{1}{\sqrt{2\omega}}
\left(\sqrt{n+1}\delta_{k,n+1}
+\sqrt{n}\delta_{k,n-1} \right) \ ,
\label{e.pgas1}
\ee
then we have:
\be
G^{(\Phi_1\rightarrow \Phi_1)}_{\beta}(t) 
= \frac{1}{\omega}\sum_{n \ge 1}n 
{\rm e}^{-\beta\, (E_n - \frac{1}{2}\Delta_n)} 
{\rm cosh}\left[\left(\frac{\beta}{2} - t\right) \Delta_n\right]/\sum_{n \ge 0}
{\rm e}^{-\beta\, E_n}
\label{e.pgas2}
\ee
For the unperturbed oscillator ($\epsilon = 0$):
\be
E_n = \left(n+\frac{1}{2}\right)\omega, \ \ \Delta_n \equiv E_n - E_{n-1}=\omega.
\label{e.osc1}
\ee
Then the $t-$dependence factorizes in (\ref{e.pgas2}) and we have a trivial 
effect of the temperature:
\be
G^{(\Phi_1\rightarrow \Phi_1)}_{osc.,\,\beta}(t)
= \frac{{\rm cosh}\left[\left(
\frac{\beta}{2} - t\right)\,\omega\right]}{2\,\omega \,{\rm sinh}
\left[\frac{\beta}{2}\,\omega\right]}
\towh{}{\longrightarrow}{\sss{\beta \rightarrow \infty}} {\rm e}^{-\omega\, t}.
\label{e.pgas22}
\ee
If we turn on the interaction the levels are no longer equidistant
(the effect of adding one more meson depends on the total number of mesons
present in the state) 
and the $t-$dependence is non-trivially affected by temperature.
We write:
\be
\Delta_1 = {\tilde {\omega}},\ \ \Delta_{n \geq 2} = {\tilde {\omega}} - 
\epsilon\, 
\lambda_n ,
\label{e.pgas3}
\ee
then to first order in $\epsilon$ (weakly interacting gas)
\be
G^{(\Phi_1\rightarrow \Phi_1)}_{w.i.g.,\,\beta}(t) \propto {\rm cosh}\left[
\left(\frac{\beta}{2} - t\right)\,{\tilde {\omega}}^{({\beta})}\right]
+ {\cal O}(\epsilon^2)
\label{e.pgas4}
\ee
with
\be
{\tilde {\omega}}^{({\beta})}
 = {\tilde {\omega}} - \epsilon \,\frac{\sum_{n \ge 1}\lambda_{n+1}\,
 (n+1)\,
\,{\rm e}^{-\beta\,E_n}}{\sum_{n \ge 0}(n+1)
\,{\rm e}^{-\beta\,E_n}}
\towh{}{\longrightarrow}{\sss { \beta \rightarrow \infty}\ {\rm or}\ 
\epsilon \rightarrow 0}
 {\tilde {\omega}} \towh{}{=}{\sss \epsilon = 0} \omega .
\label{e.pgas5}
\ee
\no {\it Notice that the above effects show up although we use the ``perfect" 
source $\Phi_1$: they represent the genuine temperature effects for an
interacting system.} From (\ref{e.pgas4}),(\ref{e.pgas5}) we see that
 as long as the interaction between the modes (``the mesons")
is weak we expect 
small changes which may be
{\it simulated} by a shift (and possibly a widening) of the
peak in the spectral function, defining in this way an effective mode
(\ref{e.effm}).
Large changes, on the other hand, will signal the installation of
a new regime. Then
we must try to obtain additional information by other tests. Essentially,
this is our program.  
Of course in real life
we shall not be able to obtain a ``perfect" source in the
above sense. The 
various uncertainties
inherent in our procedure will be repeatedly discussed in the course
of the paper.

If we use a ``perfect" source but a different sink 
(with non-zero projection on the source)
we reach similar expressions. To the next order in $\epsilon$, however,
 at $T>0$ 
the temperature correction to the mass will depend on the sink 
operator.  Generally therefore at $T>0$ we expect to find sink dependence 
of propagators even for ``perfect" source. This dependence 
can be seen as an
indication for the importance of  temperature effects.

\subsection{Mesonic correlators}
\label{sec:Mec}

A first attempt to optimize the mesonic operators, in the spirit 
described in the previous section, is to introduce a smearing function
$\omega(\vec{y})$, such that the zero-momentum mesonic operator reads:
\begin{equation}                                                                
\Phi_{\rm M}^{(\omega)}(t) = \sum_{\vec{z}} \sum_{\vec{y}} \omega(\vec{y})\,         
    \bar{q}(\vec{z},t){\gamma}_{\rm M} q(\vec{z}+\vec{y},t).                 
\label{eq:smr_opr}                
\end{equation}
giving rise to smeared correlators
(we shall omit the index ``$\beta$" in the
following):
\bea
G_{\rm M}^{(\omega\rightarrow \omega')}(t) &=& 
\langle {\rm Tr} \left[\Phi_{\rm M}^{(\omega')}(t) \Phi_{\rm M}^{(\omega)}(0)
\right] \rangle \label{e.corrq}\\                                           
&=&\mbox{$\sum_{\vec{z_1},\vec{z_2},\vec{y_1},\vec{y_2}}$}\,                  
\omega'(\vec{y_1})\omega(\vec{y_2})\,                              
\langle {\rm Tr} \left[S(\vec{z_2},0; \vec{z_1}, t) \gamma_{\rm M}
\gamma_5 S^{\dag}(\vec{z_2}+\vec{y_2},0; \vec{z_1}+ \vec{y_1}, t) \gamma_5
\gamma_{\rm M}^{\dag}\right]\rangle \nonumber                           
\eea
Here $\langle .\rangle$ means summation over Yang-Mills configurations.
$S$ is the quark propagator and $\gamma_{\rm M} =
\{\gamma_5,  \gamma_1, 1, \gamma_1\gamma_5\}$ for
M=$\{{\rm Ps, V, S, A}\}$ (pseudo-scalar, vector, scalar and
axial-vector, respectively). In the scalar sector,
only the connected part of the correlator is evaluated.
The Coulomb gauge is used to produce the quark propagators $S$ -- this is
of course irrelevant for the $\vec{y_1}=\vec{y_2}=0$ expectation values.
Note that generally 
we keep different smearing functions at the source and the sink.
This will allow, given a certain basis of operators $\{\Phi^a\}$, to perform
a variational analysis in order to attempt a further optimization of the
mesonic operator - for details see section 4.

 As smearing functions we shall use two different kinds,
\begin{itemize}
\item[i)]
 A ``point" source (sink):
\be
\omega_{\vec{x}}(\vec y) \propto \delta(\vec x-\vec y).
\ee
This will be mainly used to study the $\vec x$ dependence of the correlator 
\be
 G_{\rm M}^{(\omega)}(\vec x,t) \equiv
G_{\rm M}^{( \omega\rightarrow \omega_{\vec x})}(t) =
\sum_{\vec{z}} \langle \bar{q}(\vec{z}+\vec{x},t)
      {\gamma}_{\rm M} q(\vec{z},t)\, \Phi_{\rm M}^{(\omega)}(0) \rangle.
\label{eq:gencor}
\ee
at fixed $t$. $G_{\rm M}^{(\omega)}(\vec x,t)$ can be interpreted as ``Coulomb 
gauge wave-function", it
indicates the spatial correlation between quark and anti-quark.
\item[ii)]
 The convolution:
\bea
\omega_{a b}(\vec{y})&=&\mbox{$\sum_{\vec{v}}$}\,
\omega_a(\vec{v})\,\omega_b(\vec v+\vec{y}) \label{e.sourm}
\eea
which is equivalent to using smeared {\em quark} and {\em antiquark fields}
with smearing functions $\omega_b$ and $\omega_a$ respectively.
We use here three kinds of {\em quark} smearing functions:
\bea
 \omega_{p}(\vec{y}) &\propto& \delta(\vec{y})\ \ {\rm (}``point"{\rm )}
\label{e.sop}\\ 
 \omega_{e}(\vec{y}) &\propto& {\rm exp}(-\,a\,|\vec y|^p)\ \ {\rm (}``exp."{\rm )}
\label{e.soe}\\
\omega_{w}(\vec{y}) &\propto& 1\ \ {\rm (}``wall"{\rm )}
\label{e.sow}
\eea
In tuning the exponential source $exp$ in (\ref{e.soe}) we shall use
the parameters $a,p$ from the observed
dependence on ${\vec x}$ at large $t$ of the temporal Ps wave function with {\it point-point}
source at $T \simeq 0$,  $G_{\rm M}^{(pp)}(\vec{x},t)$ - see \ref{eq:gencor}.
The mesonic operator ``exponentially"  smeared both at the quark and the 
antiquark corresponds to a mesonic source in the relative $q$-$\bar{q}$ distance
given by the convolution (\ref{e.sourm}).
Therefore the $exp$-$exp$ smearing with $a,p$ from the wave function
implies a meson source typically {\it wider} than
the measured wave function $G_{\rm M}^{(pp)}(\vec{x},t)$, $t\gg 1$.

\end{itemize}

The $t$ dependence of the temporal propagators 
$G_{\rm M}^{(\omega\rightarrow \omega')}(t)$ depends on the spectral functions.
On a {\it periodic} lattice the contribution of a pole
in the mesonic spectral function to the $t$-propagator
is $\propto{\rm cosh}[M(t-N_{\tau}/2)]$ 
(this $M$ 
is therefore called ``pole-mass").\footnote{More
precisely, the relation between the slope parameter in  cosh, say ${\tilde M}$
and the position of the pole, $M$ is 
\be
M=\xi\sqrt{2({\rm cosh}\left({\tilde M}/\xi) - 1\right)}
\label{e.mcutb}
\ee
(and correspondingly
\be
M= {\rm ln}({\tilde M}/\xi+1)
\label{e.mcutf}
\ee
for fermionic propagators)\cite{HNS93a}. 
In the following we shall neglect these 
corrections, since they remain below the other uncertainties of our data.}
A broad structure or the admixture of
excited states leads to a superposition of such terms. Fitting
a given $t$-propagator by  cosh$(m(t)(t-N_{\tau}/2))$ at pairs of
points $t,t+1$
\begin{equation}
 \frac{G(t)}{G(t+1)}
 = \frac{\mbox{cosh}\left[m^{(\tau)}_{eff}(t)\,(N_{\tau}/2 -t)\right]}
        {\mbox{cosh}\left[m^{(\tau)}_{eff}(t)\,(N_{\tau}/2 -t-1)\right]}
\label{e.efem}
\end{equation}
defines an ``effective mass" $m_{eff}(t)$, which is a constant
if the spectral function has only one, narrow peak.
The effective mass is a rather sensitive observable
which shows effects of the source dependence,
widening of the spectral function or existence of excited states,
without, however, allowing to differentiate among them. To the
extent that the effective mass reaches a plateau and permits to define a
``temporal mass" $m^{(\tau)}$ at $T>0$, the latter
connects directly to the (pole) mass of the mesons below $T_c$, while above
$T_c$ it will presumably help  analyze the dominant low energy
structure in the frame of our strategy.
By contrast, using {\it spatial
propagators} we shall extract the
``screening mass" $m^{(\sigma)}$. 

Errors are estimated by the single elimination Jackknife method,
unless otherwise notified. For details see section \ref{s.sources}.
For various comparisons we shall also use in 
(\ref{e.corrq}), instead  of the  quark 
propagators $S$ measured in each MC configurations, free quark propagators,
defining in this way ``free" mesons.

\section{Lattice setup}
\label{sec:Setup}

In this section we describe the preparation of
 the lattice on which mesonic correlators
at zero and finite temperature are calculated.

\subsection{Anisotropic lattice}

We use  anisotropic lattices on which the spatial
and the temporal lattice spacings are different:
$a_{\sigma} \neq a_{\tau}$  \cite{Kar82}.
The simplest generalization of commonly used Wilson actions
for gauge and quark fields is obtained as follows.

For the gauge field action,
\begin{equation}
S_{G} =
   \frac{\beta}{\gamma_G} \sum_{x, i < j \leq 3}
        \left( 1 - \frac{1}{3} \hbox{ReTr}\, U_{ij}(x) \right)
 + \beta\, \gamma_G \sum_{x, i \leq 3}
        \left( 1 - \frac{1}{3} \hbox{ReTr}\, U_{i4}(x) \right)
\end{equation}
where
\begin{equation}
 U_{\mu\nu}(x) = U_{\mu}(x) U_{\nu}(x+\hat{\mu})
              U_{\mu}^{\dag}(x+\hat{\nu})U_{\nu}^{\dag}(x) 
\end{equation}
The bare anisotropy parameter $\gamma_G$ controls the 
ratio of spatial and temporal lattice spacings.
For the fermion,
\begin{equation}
 S_{F} = \frac{1}{2 \kappa_{\sigma}} \sum_{x,y}\,
  \bar{q}(x)\, K(x,y)\, q(y),
\end{equation}
\begin{eqnarray}
 K(x,y) = \delta_{x,y}
  &-& \kappa_{\sigma}
    \sum_{i=1}^3 \left[  (1-\gamma_i)\, U_i(x)\,\delta_{x+\hat{i},y}
            + (1+\gamma_i)\, U_i(x-\hat{i})\,\delta_{x-\hat{i},y} \right]
  \nonumber \\
  &-& \kappa_{\sigma} \,\gamma_F
           \left[  (1-\gamma_4)\, U_4(x)\,\delta_{x+\hat{4},y}
            + (1+\gamma_4)\, U_4(x-\hat{4})\,\delta_{x-\hat{4},y} \right],
\end{eqnarray}
where $\kappa_{\sigma}$ is the spatial hopping parameter and $\gamma_F$
is the bare anisotropy for fermions.\footnote{Note that the 
Wilson term corresponding to this 
ansatz has not a Lorentz invariant naive continuum limit. Since this
is an irrelevant operator this feature should not affect our results,
in fact this ansatz is more efficient in damping the additional
fermionic modes. This may be different for quantities where the Wilson
term acts marginally.}
For later convenience, we define $\kappa$ as
\begin{equation}
 \frac{1}{\kappa} \equiv \frac{1}{\kappa} - 2\,(\gamma_F -1)
   = 2 \,(m_0 +4 ),
\label{eq:def_kappa}
\end{equation}
where $m_0$ is the bare quark mass parameter in units of the spatial
lattice spacing.
At a later stage, we shall carry out the chiral extrapolation in this
$1/\kappa$.

The actual anisotropy $\xi$ is defined using certain correlators, $F$,
containing gauge and fermion fields.
In general, $\gamma_G$ and $\gamma_F$ are different from 
$\xi$ because of the interaction \cite{Kar82, Bur88}.
To obtain the desired value of $\xi$, one needs to tune the values of
$\gamma_G$ and $\gamma_F$ by requiring the isotropy of correlators
in physical units:
\begin{equation}
 F_{\sigma}(z) = F_{\tau}(t = \xi\, z),
\label{e.phiso}
\end{equation}
where t and z are understood in the corresponding lattice units, 
$a_{\tau}$ and
$a_{\sigma}$ respectively.
This renormalization procedure is called  ``calibration''.
Since we compare the temporal and the screening masses,
it is  important to obtain $\xi$ as precisely as possible,
and verify that these two kinds of mass coincide at $T=0$.

In the case of dynamical quarks, a precise calibration requires a
large effort, since we generally have four bare parameters 
($\beta$, $\gamma_G$, $\kappa_{\sigma}$ and $\gamma_F$) and only
three physical parameters $\xi$, $a_{\sigma}$ and $m_q$ which can be varied, 
therefore the condition of physical isotropy implies
a non-trivial constraint among the bare parameters \cite{HNS93b}.
In a quenched simulation, however, the situation is much simpler.
Here one generates gauge field configurations using certain values
of $\beta$ and  $\gamma_G$ and reads $\xi$ from gluonic
quantities.
Then the fermionic parameters are determined such that they give the
desired quark (or hadron) mass and the same $\xi$ is obtained from 
hadronic correlators.

\subsection{Lattice parameters}

We use  lattices of sizes $12^3 \times N_{\tau}$,
where $N_{\tau}=72$ ($T=0$), 20 (below $T_c$), 16 and 12 (above $T_c$),
with couplings $\beta=5.68$, $\gamma_G=4.0$,
in the quenched approximation.
\footnote{The lattice described in this paper corresponds to {\sl Set-B}
data in our earlier reports \cite{TARO99}.}
Configurations are generated with the pseudo-heat bath algorithm 
with 20000 thermalization sweeps, the configurations being
separated by 2000 sweeps.
In most cases, 60 configurations are used, except for the calibration.
The gauge field is fixed to the Coulomb gauge. The calibration is described
in detail in the Appendix.
From the calibration of the gauge configurations we obtain a 
renormalized anisotropy  $\xi=5.3(1)$.
The lattice cutoffs determined from the heavy quark potential
are $a_{\sigma}^{-1} = 0.85(3)$ GeV
($a_{\sigma} \sim 0.24$ fm) and $a_{\tau}^{-1} = 4.5(2)$ GeV
($a_{\tau} \sim 0.045$ fm).

Table~\ref{tab:parameters} summarizes the quark parameters and gives the
meson masses as determined in the next section.
As a guide for the mass range we are concerned with, 
the quark mass is estimated
by a naive relation $m_q=(\kappa^{-1}-\kappa_c^{-1})/2$ 
using the critical hopping parameter $\kappa_c=0.17144(11)$. 
This simulation deals therefore with quarks in the strange quark mass region.
The boundary conditions for quark fields are set to periodic and anti-periodic
in spatial and temporal directions respectively, except for the calibration
(at $T=0$) where anti-periodic b.c. in all four directions are used.

In the following we always display  dimensionless quantities, that is,
they are understood as given in the corresponding lattice spacings or
their inverses.
Since we have two lattice spacings,
$a_{\sigma}$ and $a_{\tau}$, when we shall compare different quantities 
we shall use the translation $a_{\sigma}= \xi a_{\tau}$, i.e., the
numbers featured can be understood as given in units of 
$a_{\tau}$ ($a_{\tau}^{-1}$).

\begin{table}[tb]
\begin{center}
\begin{tabular}{ccccccc}
\hline \hline
$\kappa_{\sigma}$ & $\gamma_F$ &  $N_{conf}$ & $\kappa$ &  $m_q$  &
              $m_{Ps}$ & $m_{V}$ \\
\hline
0.081 & 4.05(2) & 20 & 0.1601 & 0.0389 & 0.1777( 8) & 0.1962(10) \\
0.084 & 3.89(2) & 20 & 0.1633 & 0.0276 & 0.1493( 9) & 0.1747(12) \\
0.086 & 3.78(2) & 30 & 0.1648 & 0.0223 & 0.1341(10) & 0.1644(13) \\
\hline \hline
\end{tabular}
\end{center}
\caption{Quark parameters.
$\gamma_F$ is determined by the calibration using first
$N_{conf}$ configurations.
The error of $\kappa$ and $m_q$ is not estimated ($\gamma_F$ is fixed
in successive calculation).
The tabulated meson masses are the values obtained with smeared correlators
described in the next section. All masses are given in units of
$a_{\tau}^{-1}$.}
\label{tab:parameters}
\end{table}

\subsection{Finite temperature}

In \cite{TARO96} we have measured the Polyakov loop susceptibility
as a function of $\gamma_G$ for several values of $N_{\tau}$.
At $N_{\tau}=18$, the peak was found between $\gamma=3.9$ and
4.0, which means on our $\gamma=4$ lattice that
$N_t=18$ is very close to and just above $T_c$.
The estimated temperature for our values of $N_{\tau}$ are therefore 
$T\simeq 0$, $0.93\, T_c$, $1.15\, T_c$
and $1.5 \,T_c$ for $N_{\tau}=72$, $20$, $16$ and $12$ respectively. 

We found that the configurations above $T_c$ stay in a single Polyakov loop
sector during the whole updating history, and that the behavior of meson
correlators strongly depends on the sector. The hadronic
correlators feel the deconfining
transition if they are taken in the real sector, but they appear not to
``notice" it if they are taken on configurations in one of
the complex sectors  \cite{TARO99}.
With dynamical quarks, the $Z_3$ center symmetry is explicitly broken and 
the Polyakov loop  prefers to stay on the real axis.
Since we regard the quenched lattice as an approximation to the dynamical 
lattice, we restrict our simulation to the case with real Polyakov loop
sector.

\section{Zero temperature analysis.}
\label{sec:zerotemp}

In the context of our strategy 
the analysis of the zero temperature correlators serves as  foundation 
for the analysis at
 $T>0$. This is also a  good opportunity to describe in detail 
our procedure. Here we obtain
 the meson masses and the wave function which is used to
smear the source and sink operators.

\subsection*{Smearing of mesonic operators}
\label{s.sources}

To fix the {\it exponential} smearing function we measure the wave function
at zero temperature, given by (\ref{eq:gencor}) with point smearing 
$\omega_p(\vec{y})=\delta(\vec{y})$ both for the quark and anti-quark
at the source:
and define 
\be
\omega_e(\vec{y}) = G_{\rm M}^{(pp)}(\vec{y},t)/G_{\rm M}^{(pp)}(\vec{0},t)
 \large|_{t\gg 1} 
 = \exp(-\,a\, y^p),
\ee
where $a$, $p$ are fitting parameters.
The fitted values of $a$ and $p$ of the Ps meson wave function are
listed in Table~\ref{tab:mass72}.

\begin{figure}[tb]
\vspace*{-0.5cm} 
\center{
\leavevmode\psfig{file=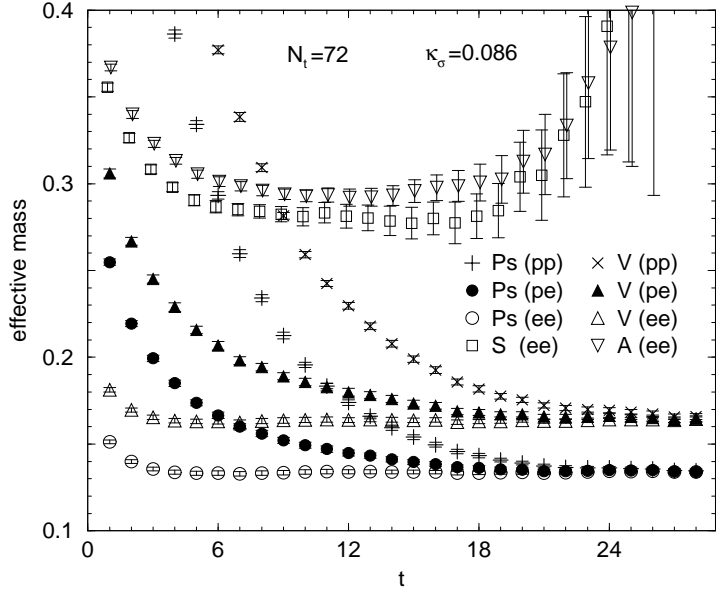,width=\figwidth}}
\vspace{-0.5cm} \\
\caption{
The effective masses of correlators with various source smearing
functions and the point sink for $\kappa_{\sigma}=0.086$, at $N_{\tau}=72$.}
\label{fig:ep72}
\end{figure}

\begin{figure}[tb]
\center{
\leavevmode\psfig{file=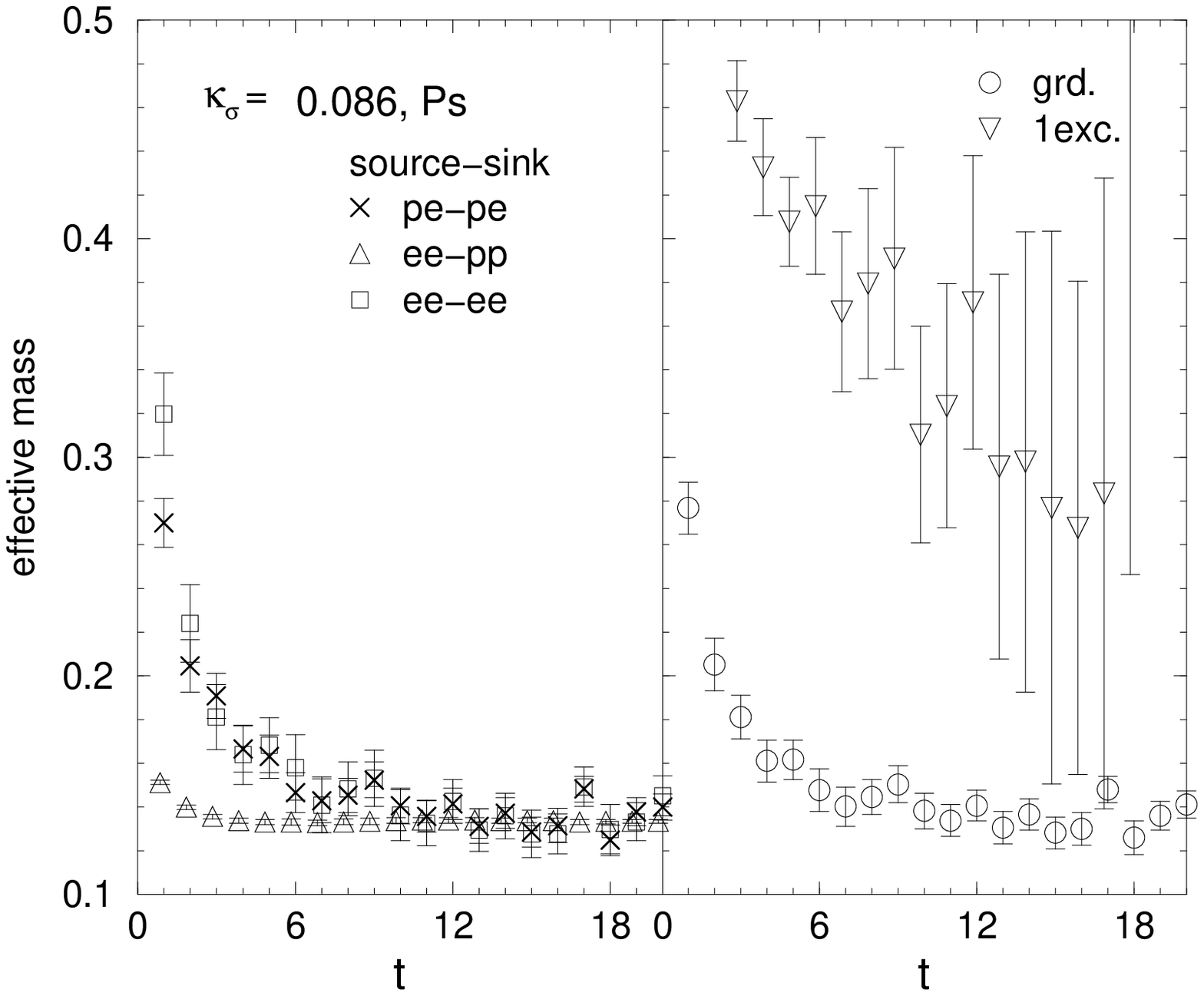,width=\figwidthb}
\leavevmode\psfig{file=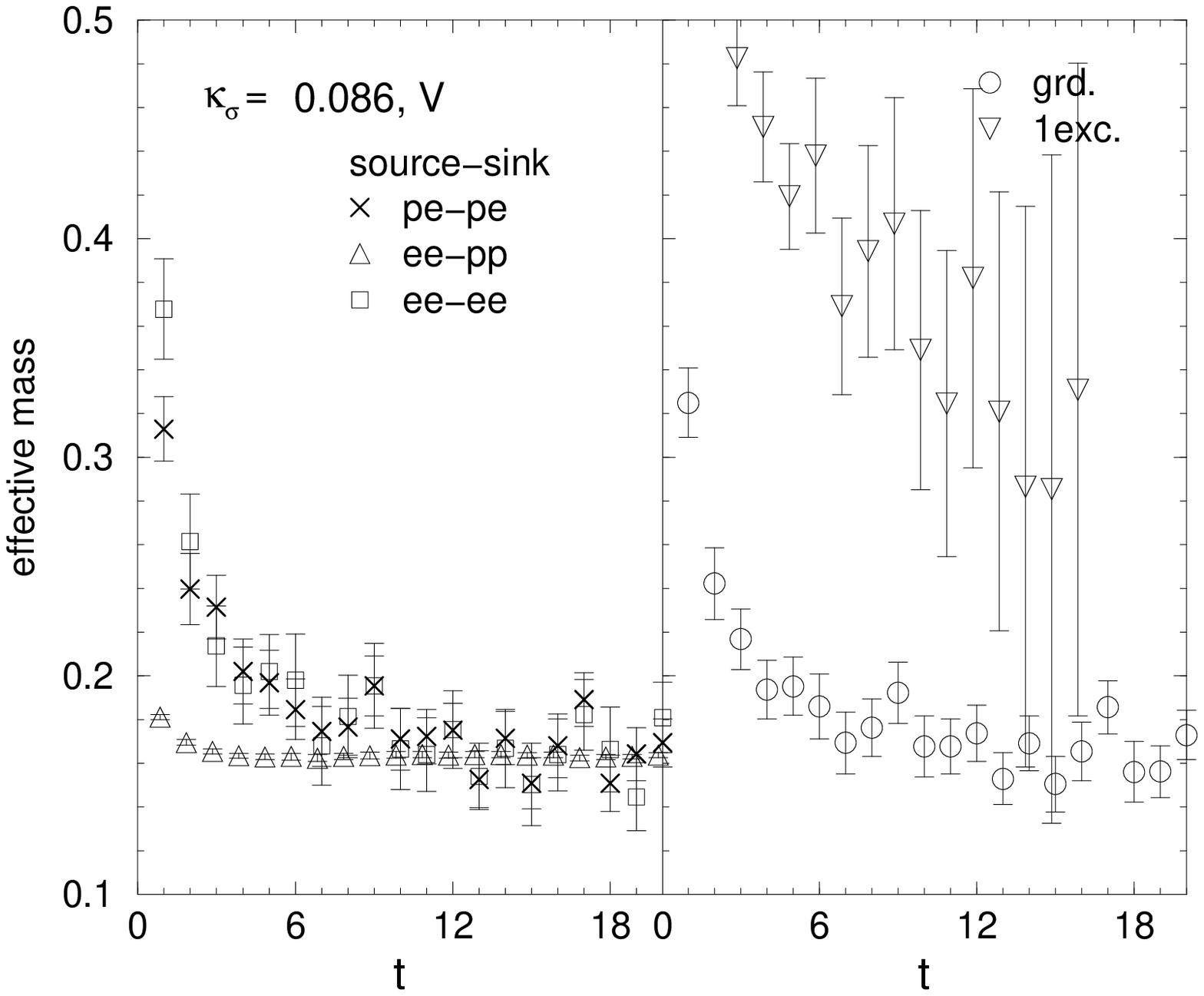,width=\figwidthb}}
\caption{
Effective masses of Ps and V correlators for $\kappa_{\sigma}=0.086$
at $N_{\tau}=72$.
In the left plots for each figure, propagators with various source and sink 
smearings
are compared to observe the dependence on smearing.
The right plots show the result of the variational analysis:
the effective masses of the ``ground"
 and the ``first excited" states are
displayed.}
\label{fig:var_ep72}
\end{figure}

To extract the effective mass from the correlators following eq. (\ref{e.efem})
we have several possibilities depending on the choice of mesonic operators
$\Phi_M^{(\omega)}(t)$. We call $m_{\rm eff}^{(\omega\rightarrow \omega')}(t)$ 
the effective mass
extracted from correlators smeared with $\omega$ and $\omega'$ at the source and the sink
respectively.
In figure~\ref{fig:ep72} we display the effective masses
for $\kappa_{\sigma}=
0.086$.  In all cases, the sink $\omega'$ is a $point$-$point$ operator and we show 
three choices of $\omega$ at the source: $point$-$point$, $point$-$exp$ and
$exp$-$exp$ ($pp$, $pe$ and $ee$ respectively in what follows).
For S and A channels, only the correlators  with $exp$-$exp$ source
are shown, since with other sources statistical fluctuations are so
large that no clear plateaus are  found.
It is clear from the figure that the effective masses from different
operators converge to the same value at large $t$, the worst behavior being 
observed for the non-smeared $pp$-$pp$ 
correlator. Considerable improvement is observed for the masses extracted 
from smeared operators. Of them $m_{\rm eff}^{(ee\rightarrow pp)}(t)$ is 
the one that converges
most rapidly to a plateau, though it increases slightly at early stages, 
which is due to the fact that the source is slightly too wide, as discussed 
in section \ref{sec:Mec}. 
The amount of optimization achieved by the exponentially smeared operator
has to be analyzed through the diagonal correlator $\langle \Phi_M^{ee}
(t) \Phi_M^{ee} (0) \rangle$ which
is a sum of positive contributions from the different states -- see
 (\ref{e.pt0}), (\ref{e.ts0}). 
In Fig.~\ref{fig:var_ep72} we show, in addition to the off-diagonal $ee$-$pp$
effective mass,  the masses extracted from diagonal $ee$-$ee$ and $pe$-$pe$ correlators.
 They both show a very similar behavior and do not reach a plateau
up to $t\sim 10$, where they merge with the $ee$-$pp$ result. 
This is an indication that the ``good" behavior
of $m_{\rm eff}^{(ee\rightarrow pp)}(t)$ is partly due to an ``accidental" 
cancellation of contributions from higher excited states which
in a non-diagonal correlator may have alternating signs. Therefore also 
the $m_{\rm eff}(t)$ extracted from such correlators is no longer bounded from below
by the meson mass.

We have tried to improve the mesonic operator by performing a variational 
analysis in the basis of operators $\{\Phi^{pp},\Phi^{pe},\Phi^{ee}\}$.
This amounts to a diagonalization providing the best approximation to the 
ground and two first excited states within the space of operators we have used.
The result of the diagonalization is also shown on Fig.~\ref{fig:var_ep72}
where the effective masses
of the ``ground" and the ``first excited" states are displayed
(the ``second excited" state suffers from large fluctuations).
As can be seen, within this basis of operators no improvement
is obtained. The ground state effective mass is very similar to the
$ee$-$ee$ and $pe$-$pe$ ones, probably an indication that the basis
of operators used is too correlated to provide any further improvement.

Summarizing the observations of the source dependence and the variational
analysis at $T=0$, within the basis of three operators we have at present, 
the analysis does not achieve further optimization of the correlators. 
Since the effective masses extracted from all sources approach  the 
$ee$-$pp$ one and the latter reaches earlier a plateau 
we shall use the $ee$-$pp$ correlator  for the coming discussion.
One should however keep in mind that such correlator is not really
optimized in the sense of been constructed from a sufficiently optimized 
meson source at $T=0$. There is a priori no guarantee that the cancellation
taking place at $T=0$ will still remain at $T>0$. We will use the departure
of $m_{\rm eff}^{(ee\rightarrow pp)}$ from flatness as an indication
that temperature effects start to become relevant.
To control the uncertainties introduced by
this choice we keep measuring correlators with the different types of operators,
to investigate the effective masses source dependence also at $T>0$. 
As long as the effective 
masses extracted from different sources converge to the same value,
temperature effects will be small and the extraction of the meson mass
from the $ee$-$pp$ correlator will be safe. When the source dependence starts
to be important we will rely in addition on other type of analysis, 
like the study of the wave function and the comparison with the free 
quark case.

\subsubsection*{Spectroscopy at T=0}

We briefly summarize here the meson spectroscopy.
We extract the masses at $T=0$ from the $ee$-$pp$ propagators for simplicity
(see the discussion above).
For the pseudo-scalar and vector meson, the masses are extracted
from a fit to a single exponential in the range $t=27-36$
(where the three types of correlators coincide; although the  $ee$-$pp$
propagator reaches a plateau much earlier, precision was not lost
by this limitation).
For the S and A channels, the statistical errors are much larger than
for Ps and V, therefore we adopted the fitting region $t=12-20$ for the 
former.
As was already mentioned, only the connected part of the scalar channel
is evaluated, hence the result is only useful for a comparison with the finite
temperature results.
The  values obtained are listed in Table~\ref{tab:mass72}.
These values are consistent with the result of the variational analysis,
where we extract masses from the fitting region $t=12-16$.

Masses are extrapolated to the chiral limit with
the definition of $\kappa$ in eq. (\ref{eq:def_kappa}).
First, the pseudo-scalar meson mass squared is extrapolated
linearly in $1/\kappa$ to determine $1/\kappa_c$ at which
the Ps meson mass vanishes. This gives $\kappa_c=0.17144(11)$.
Then the other meson masses are extrapolated linearly to $1/\kappa_c$.
The results of the extrapolation are also listed in Table~\ref{tab:mass72}.

Comparing with the physical value of the $\rho$ meson mass,
the vector meson mass at $\kappa_c$ defines the lattice cutoff as
$a_{\tau}^{-1}(m_{\rho})\sim 5.7$ GeV.
This value is about 27 \% larger than $a_{\tau}^{-1}=4.5$ GeV
from the string tension.
The discrepancy is consistent with results on isotropic lattices,
and  is mainly explained as an ${\cal O}(a)$ effect.

\begin{table}[tb]
\begin{center}
\begin{tabular}{cccccccc}
\hline \hline
$\kappa_s$ & $\gamma_F$ & $a$ & $p$ &
                      $m_{Ps}$ & $m_{V}$ & $m_{S}$ & $m_{A}$ \\
\hline
0.081& 4.05& 0.3785(33)& 1.289(8)& 0.1777( 8)& 0.1962(10)& 0.302(5)& 0.314(5)\\
0.084& 3.89& 0.3797(31)& 1.277(8)& 0.1493( 9)& 0.1747(12)& 0.285(7)& 0.300(6)\\
0.086& 3.78& 0.3800(25)& 1.263(8)& 0.1341(10)& 0.1644(13)& 0.280(9)& 0.296(6)\\
\multicolumn{2}{c}{$\kappa_c$} & - & - 
                             &     -      & 0.1225(16)& 0.248(13)& 0.269(9)\\
\hline \hline
\end{tabular}
\end{center}
\caption{
The meson masses at zero temperature in units $a_{\tau}^{-1}$.
For the scalar channel, only the connected part is evaluated.
$a$ and $p$ are the fitted parameters of the observed Ps wave function, 
and they are used to smear the quark source.}
\label{tab:mass72}
\end{table}

\section{Non-zero temperature}
\label{sec:FT}

In this section, we study how the temperature changes the
meson correlators.
First we shall compare propagators with various source and
sink operators. Then we shall 
discuss in more detail what we can learn from the temperature behavior of the
effective masses. For a further insight in the 
temperature effects on the meson 
correlators we study the $t$-dependence of the wave functions.
As a result we find that the two quarks tend to stay together
even in the deconfining phase (at least for Euclidean time scales $\sim 1/T$).
Finally we study the temperature behavior of temporal (``pole") and 
spatial (``screening") masses which could be associated
with the putative quasi-particle (resonance?) states suggested by the first
steps of the analysis.

To disentangle perturbative from non-perturbative effects we shall repeatedly
compare the measured (``full") meson correlators with ``free" meson correlators made out
of unbound quark propagators. For the latter we just use a free quark ansatz
and only allow the quark mass to vary with the temperature. This means 
that we consider quark-antiquark correlation functions in the 
corresponding mesonic channels in lowest order perturbation
theory but with a temperature dependent quark-mass. As has been shown 
by a more involved analysis, including further
 thermal effects in the HTL approximation does
not significantly change the result \cite{fri00}.  

\subsection{Source dependence of the propagators at non-zero temperature}
\label{subsec:FTcorr}

\subsubsection*{$N_{\tau}=20$ (below $T_c$)}

\begin{figure}[tb]
\center{
\leavevmode\psfig{file=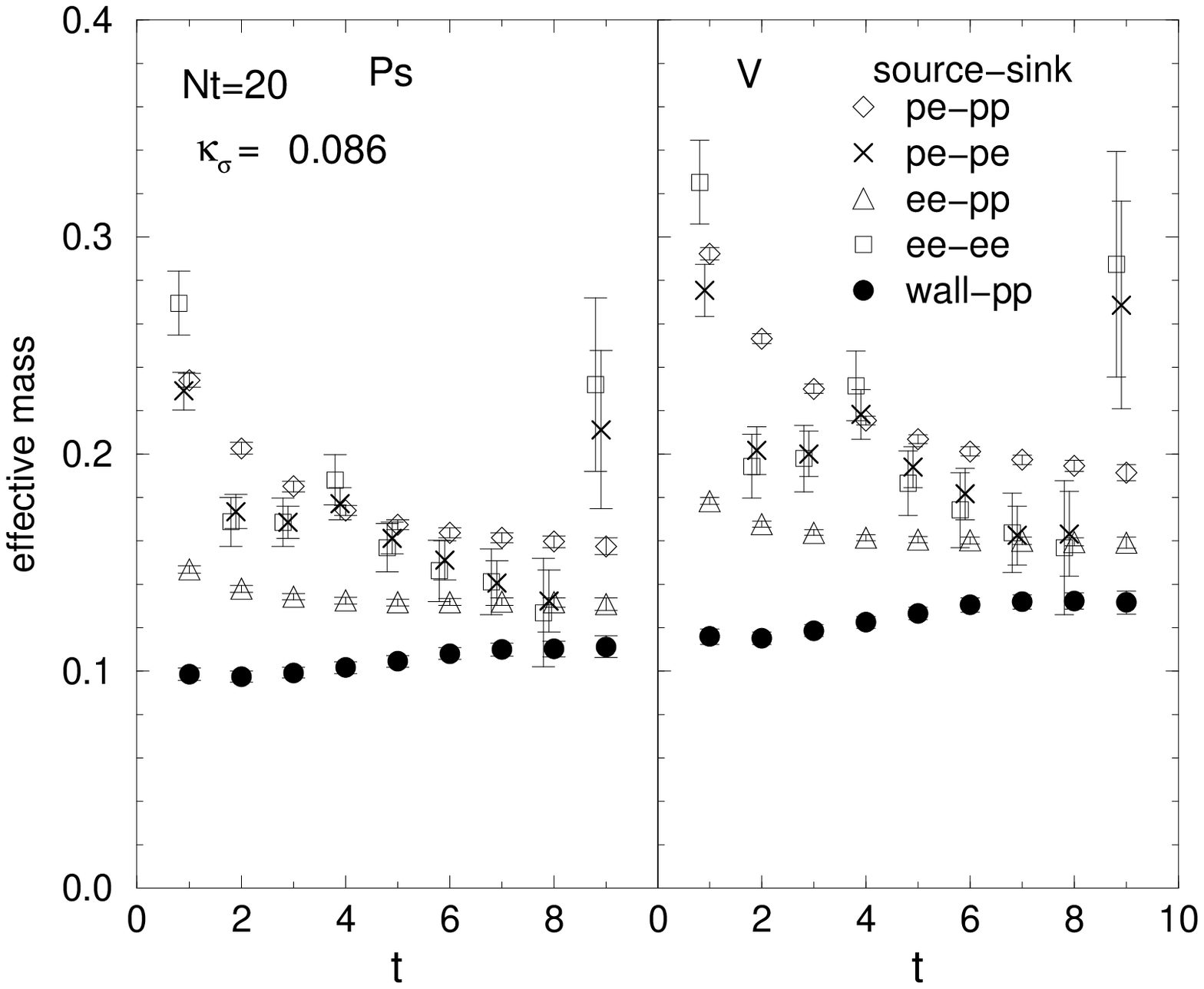,width=\figwidthb}
\leavevmode\psfig{file=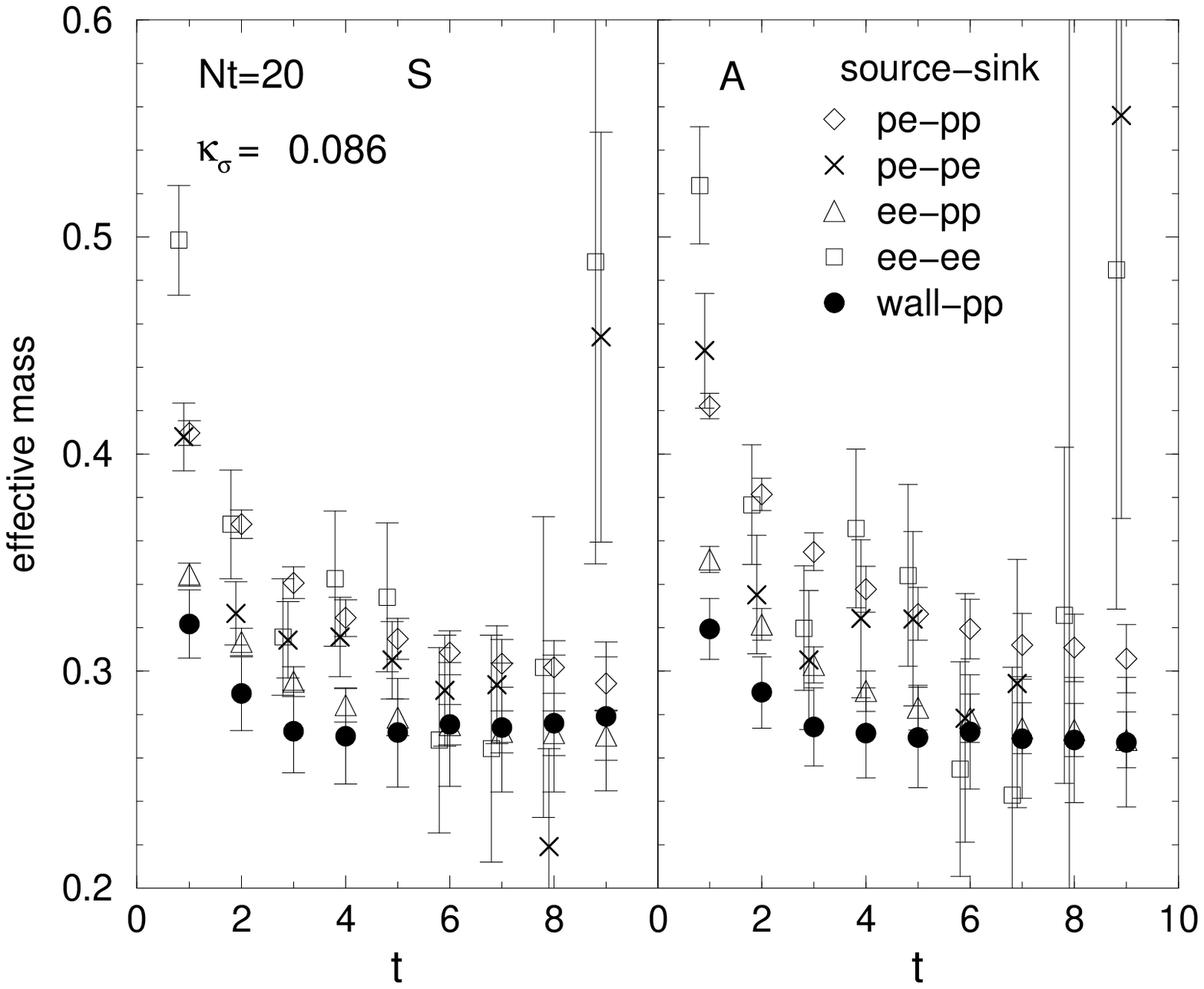,width=\figwidthb}}
\caption{
Source dependence of effective masses 
at $N_{\tau}=20$ for $\kappa_{\sigma}=0.086$, all channels.}
\label{fig:ep_20}
\end{figure}

Let us start with $N_{\tau}=20$.
The system is in the confining phase, hence we expect the hadronic spectral
function to still have narrow peaks corresponding to the bound states.
Indeed the situation is very similar to the $T=0$ case.
Fig.~\ref{fig:ep_20} compares the effective masses with several choices 
of source and sink smearing functions (we have also measured here the
effective mass with $wall$ source).
For the Ps and V channels, the $ee$-$pp$
correlator appears most flat showing, as for $T=0$, a rather clear
plateau. The two diagonal correlators, smeared both at the source and the 
sink, have strongest contributions from excited states but merge with 
the $ee$-$pp$ result at about $t\sim 7$.
(Since the smeared sink suffers from large fluctuations
the discrepancy of the effective masses at $t=9$ of the sink
smeared propagators and the propagators with point sink is
probably the result of insufficient statistics.)
Here again a variational analysis does not provide any improvement on the
$ee$-$ee$ and $pe$-$pe$ results.

The convergence of the effective masses from non-diagonal
correlators with point-point sink has not taken place yet at $t\sim 
N_t/2$. This could in principle be a first signal of temperature effects
but notice that the difference of effective masses at this value of $t$
is of the same order as  that observed at $T=0$ at the same $t$ slice.

In the S and A channels the statistical fluctuation of the effective
masses are much larger than for Ps and V.
In these channels, the effective masses from $wall$-$pp$
and $ee$-$pp$  correlators coincide.
Again we observed large fluctuations at large $t$ for the sink smeared
propagators.

Like for $T=0$, at $N_{\tau}=20$ we conclude that the most reliable 
estimate of the meson
masses is again obtained from the $ee$-$pp$ propagators.
The extracted masses are discussed in subsection~\ref{subsec:tdep}.

\begin{figure}[tb]
\center{
\leavevmode\psfig{file=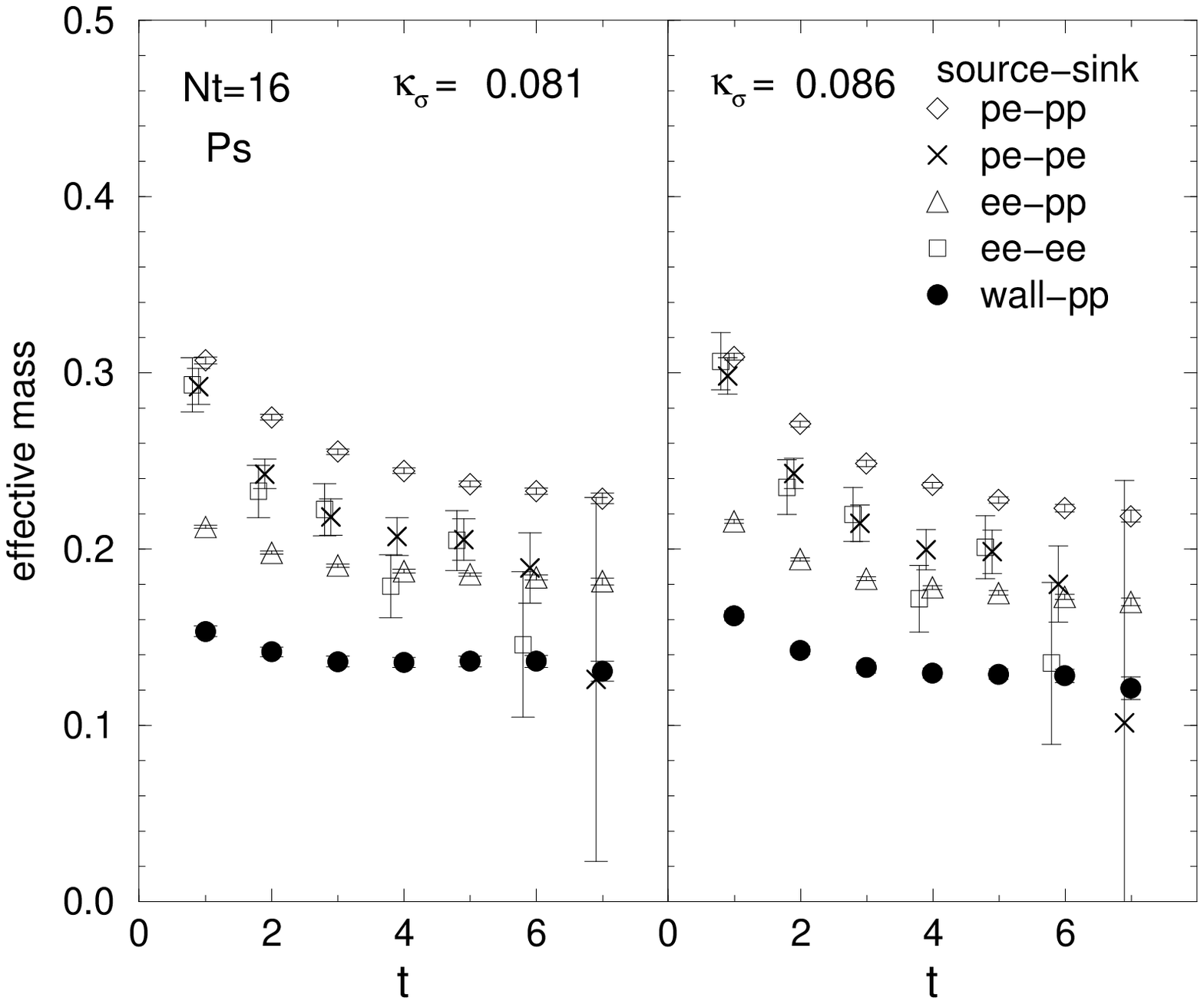,width=\figwidthb}
\leavevmode\psfig{file=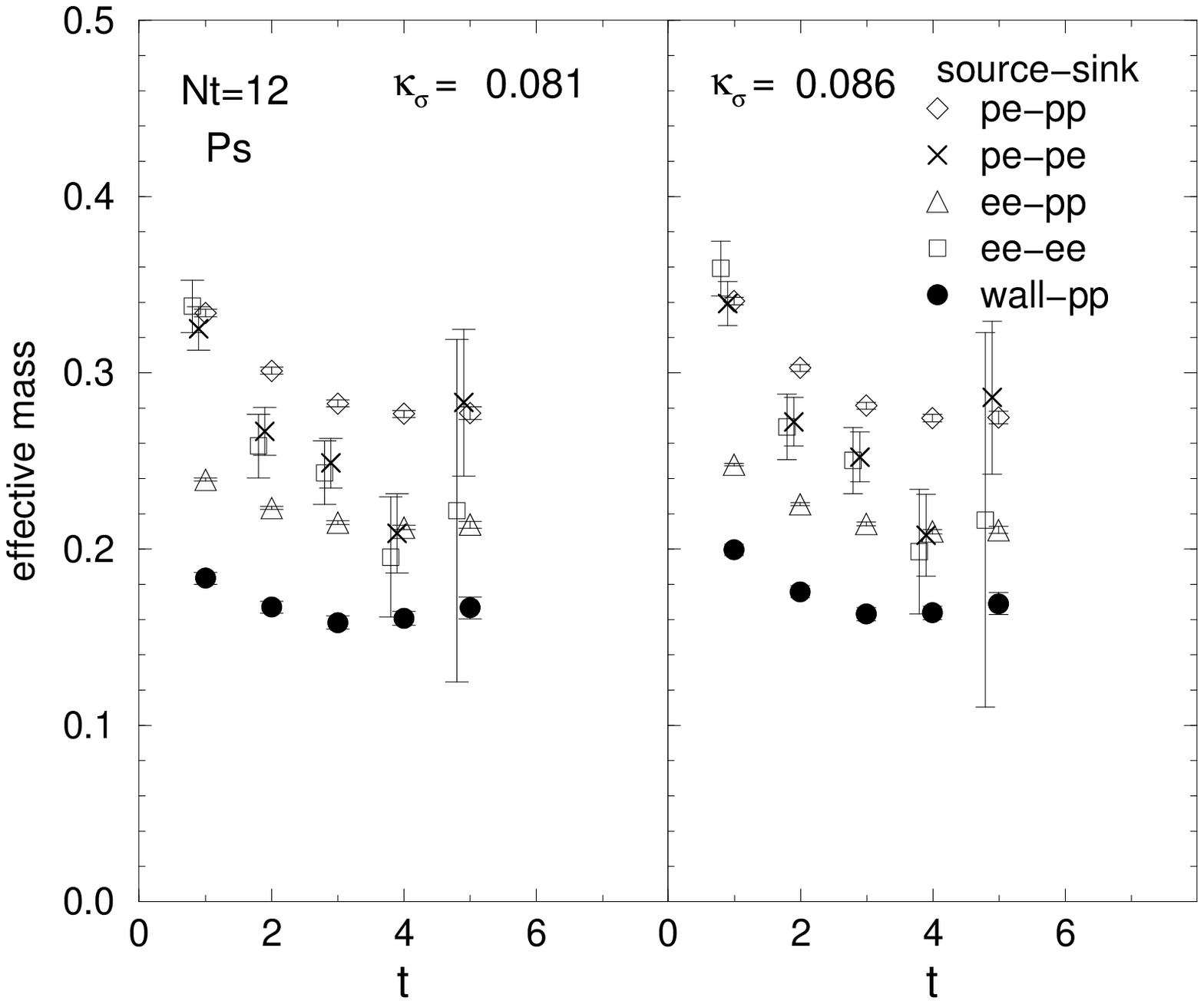,width=\figwidthb}}
\caption{
Source dependence of Ps effective masses at $N_{\tau}=16$ and $12$ for $\kappa_{\sigma}=0.081$,
 $0.086$.}
\label{fig:ep_ftvar}
\end{figure}

\subsubsection*{$N_{\tau}=16$ and $12$ (above $T_c$)}

At $N_{\tau}=16$ and $N_{\tau}=12$,
the propagators are very similar between the
different channels which is a clear indication of chiral symmetry
restoration immediately above $T_c$ (see section \ref{subsec:tdep} 
for further discussion of this point).
Figure~\ref{fig:ep_ftvar} shows the effective mass dependence on the 
mesonic operator at $N_{\tau}=16$
and $N_{\tau}=12$ for the Ps channel (the  effective 
masses in the other channels are similar).
To see the quark mass dependence, results for two values of $\kappa_{\sigma}$
are shown.
An interesting feature is that the $\kappa_{\sigma}$ dependence is small.

Although above $T_c$ the $ee$-$pp$ effective masses show no longer such a clear plateau 
(further discussion on this point will be done in section 5.2) they
 still seem to merge with those coming from diagonal
$ee$-$ee$ and $pe$-$pe$ correlators at about $t\sim    4$ (statistical fluctuations
do, however, not allow a quantitative estimate). There is, however, a clear difference
in the behavior of the non-diagonal effective masses here, as compared to
$T<T_c$. In particular the $wall$-$pp$ mass looks as flat as 
the $ee$-$pp$ but seems to converge to a rather different value (the difference
at $t\sim    N_t/2$ being here considerably larger that the corresponding
one at the same time slice for $N_t=20$).

The observed stronger source dependence above $T_c$ might just be an effect
of periodicity or contamination from excited states but it is peculiar
that this behavior sets in precisely at $T_c$. In view of the discussion
of section 2.1, we consider this as indication that strong 
temperature effects develop above $T_c$ 
(further discussion on this point will follow
later). We have also performed at this $N_t$'s a 
variational analysis which again just reproduced the values of the
$ee$-$ee$ and $pe$-$pe$ correlators.

\bigskip

\subsection{Effective mass of $ee$-$pp$ propagators at nonzero T}

\begin{figure}[tb]
\center{
\leavevmode\psfig{file=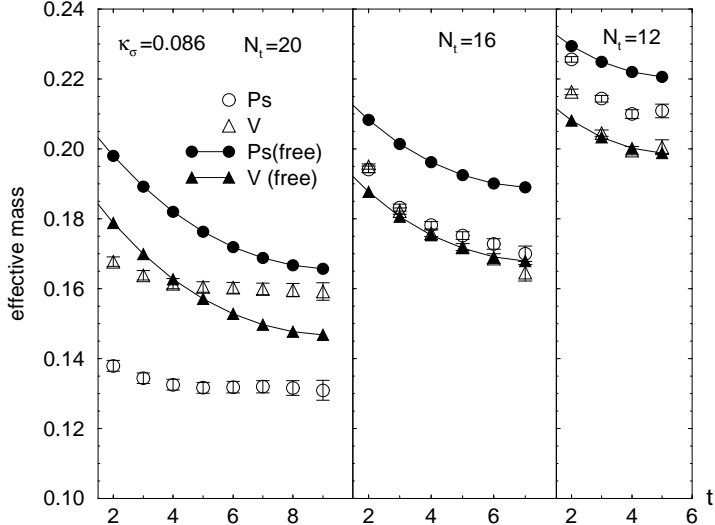,width=\figwidth}}
\caption{
Effective masses of the correlators with $exp$-$exp$ source and point sink
at finite temperature for $\kappa_{\sigma}=0.086$.
Shown are also the effective masses of the ``free" mesons,
composed of the free quark propagators.}
\label{fig:epFT}
\end{figure}

We shall first discuss the $T-$dependence of 
propagators.
Figure~\ref{fig:epFT} shows the effective masses of Ps and V
meson correlators with $ee$-$pp$ smearing
at $N_{\tau}=20$, $16$ and $12$ for $\kappa_{\sigma}=0.086$.
While at $N_{\tau}=20$ a plateau appears, at $N_{\tau}=16$ and $12$
the effective masses are no longer flat, their $t$-dependence 
increasing with the temperature. This holds already at $t$-values at
which the $N_{\tau} = 20$ data have clearly reached a plateau ($t \simeq 4$)
and indicates that we have to do here with strong temperature effects.
Due to the uncertainties related with the $ee$-$pp$ smearing it is
not possible to say whether this behavior means that the mesonic state
has become metastable above $T_c$, or it has been replaced by a collective
excitation of increasing width, or that the effects of the
contamination  with other states from the insufficiently tuned source
have become very strong. Nevertheless it is remarkable that we find a clear 
change in behavior setting in at $T_c$, although smoothly connecting  to
the behavior below $T_c$ (at least for the Ps and V propagators: the 
S and A correlators change more significantly, in accordance with
the chiral symmetry restoration). The same signal is provided by the 
observation of the source dependence (see previous section). Notice
that the contamination with other states due to the imperfection of
the source alone would  be expected to produce a rather continuous dependence
on the temperature, and not the clear difference in behavior 
observed below and above $T_c$.

Since above $T_c$ we are in the QGP phase a first thing to test is
to look for signals of the perturbative, high $T$ regime, that is for
unbound quark - antiquark propagation in the mesonic channels. 
On figure~\ref{fig:epFT} we show effective masses calculated from
 ``free" meson correlators 
constructed out of free quark propagators where only the mass is
assumed to vary with the temperature (as already noticed, a consistent HTL
calculation \cite{fri00} does not significantly  affect the result).
The free quark propagators 
are calculated with $\gamma_F=5.3$ and using the same
$exponential$ source as for the 
genuine mesons. The plotted results for the ``free quark mesons"
correspond to the assumption of a thermal
mass for the quarks:
\be
m_q^{therm} \simeq \frac{g^2}{\sqrt{6}}T \sim 0.036\frac{T}{T_c}a_{\tau}^{-1} 
\label{e.thm}
\ee
tuned such as to give a good agreement in the $\rho$ channel above $T_c$
with the measured (``full") effective masses. 
We observe that above the critical temperature 
the measured Ps and V effective masses
change their order and increase $\propto T$, feature shared with the 
``free" mesons. The inversion in the order of the masses essentially 
implies that the 
propagators are not dominated by a narrow state (compare \cite{wein}), 
however not every wide spectral function
leads to this inversion, therefore this similarity may have significance.
After we tuned the free quarks to simulate the (full) data in
the $\rho$ channel, the pseudo-scalar remains, however, 
still well below the ``free" results 
(a similar observation has been made in \cite{biel1}). 

Also the stronger source dependence observed above $T_c$ 
is a feature which the measured
(``full") meson propagators share with the free ones -- but for the latter this
is much 
more pronounced
and for particular sources quite different from the full 
mesons. For instance,
for the wall source, the effective mass of ``free" mesons
is completely flat and its value is twice the quark mass value, 
while the full meson effective
masses are significantly larger and vary with $t$.

In the analysis of the propagators we observe therefore competing
features, which hint to some contributions from ``unbound" quarks but
cannot be explained only in terms of the latter. To investigate further 
this problem 
we analyze the $t$-dependence of the wave function in the next subsection.

\subsection{Wave functions}
\label{subsec:Wavefunc}

We consider the normalized wave function at the spatial origin:
\begin{equation}
  \varphi^{(\omega)}(\vec{x},t) \equiv G_{\rm M}^{(\omega)}(\vec{x},t)/G_{\rm M}^{(\omega)}(\vec 0,t).
\label{e.nwf}
\end{equation}
We recall that these correlators are obtained in the Coulomb gauge.
If the correlator $G_{\rm M}^{(\omega)}(\vec{x},t)$ is dominated 
by a bound state $\varphi(\vec{x},t)$ should 
stabilize with increasing $t$, approaching a certain shape. If there is no
 spatial correlation among the quarks, in particular in the case of
 a ``free" correlator (constructed from free quarks), the corresponding
$\varphi(\vec{x},t)$ should become broader in $\vec{x}$
 with $t$ (or at best reproduce the source, for $m_q \rightarrow \infty$)
\footnote{For a simple illustration consider
two nonrelativistic quarks of mass $m_a$, $m_b$ and 
 individual initial
Gaussian distributions  $\psi_q(y,0)\propto {\rm exp}(-y^2/2a^2)$ and
$\psi_{\bar q}(y,0)\propto {\rm exp}(-y^2/2b^2)$, respectively,
then it is easy to see that
 the width of the distribution in the relative distance $x$,
 $\psi(x,t) =
 \int dy\, dz\, \delta (z-y-x)\, \psi_q(y,t)\, \psi_{\bar q}(z,t)$ 
(essentially, our $\varphi(x,t)$)
develops as
$\Gamma(t)^2 = \Gamma(0)^2 + (1/m_a+1/m_b)\,t$ ,
with $\Gamma(0)^2=a^2+b^2$.  Nodes in the original
 distribution may lead to occasional cancellations, but the
general picture is the same. It would be physically quite
unplausible that
 uncorrelated  propagating quarks
would tend toward a distribution in the relative coordinate 
narrower  than the
one they start with,
whatever their spectral functions might be.}. 
 We shall use different sources, $\omega$, and observe the evolution of 
$\varphi^{(\omega)}(\vec{x},t)$ with $t$. 

\begin{figure}[tb]
\center{
\leavevmode\psfig{file=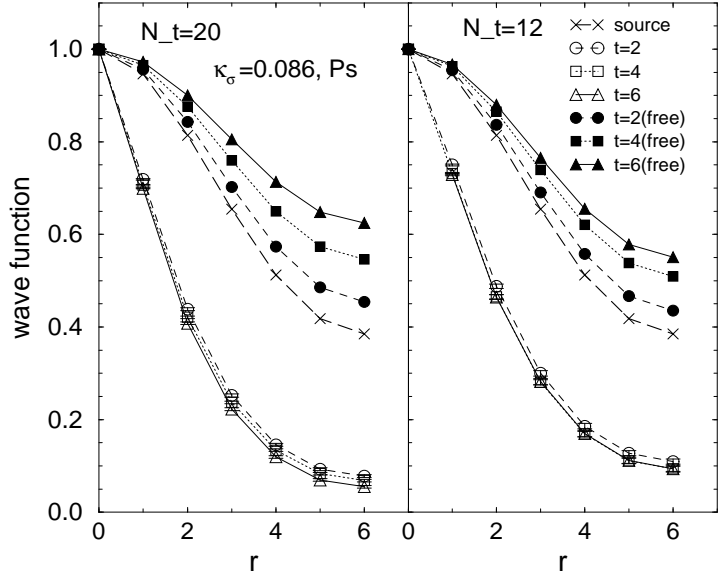,width=\figwidth}}
\caption{
$t$-dependence of the wave function $\varphi^{(ee)}(\vec{x},t)$ at
$N_{\tau}=20$ and $12$ for
$\kappa_{\sigma}=0.086$, Ps channel.
The ``free" wave function is also displayed.
Both measured (``full") and ``free" correlators are smeared with $exp$-$exp$ functions.
We represent by  crosses the convolution eq. (\ref{e.sourm})
which gives the separation distribution in the source.}
\label{fig:wft}
\end{figure}

\begin{figure}[tb]
\center{
\leavevmode\psfig{file=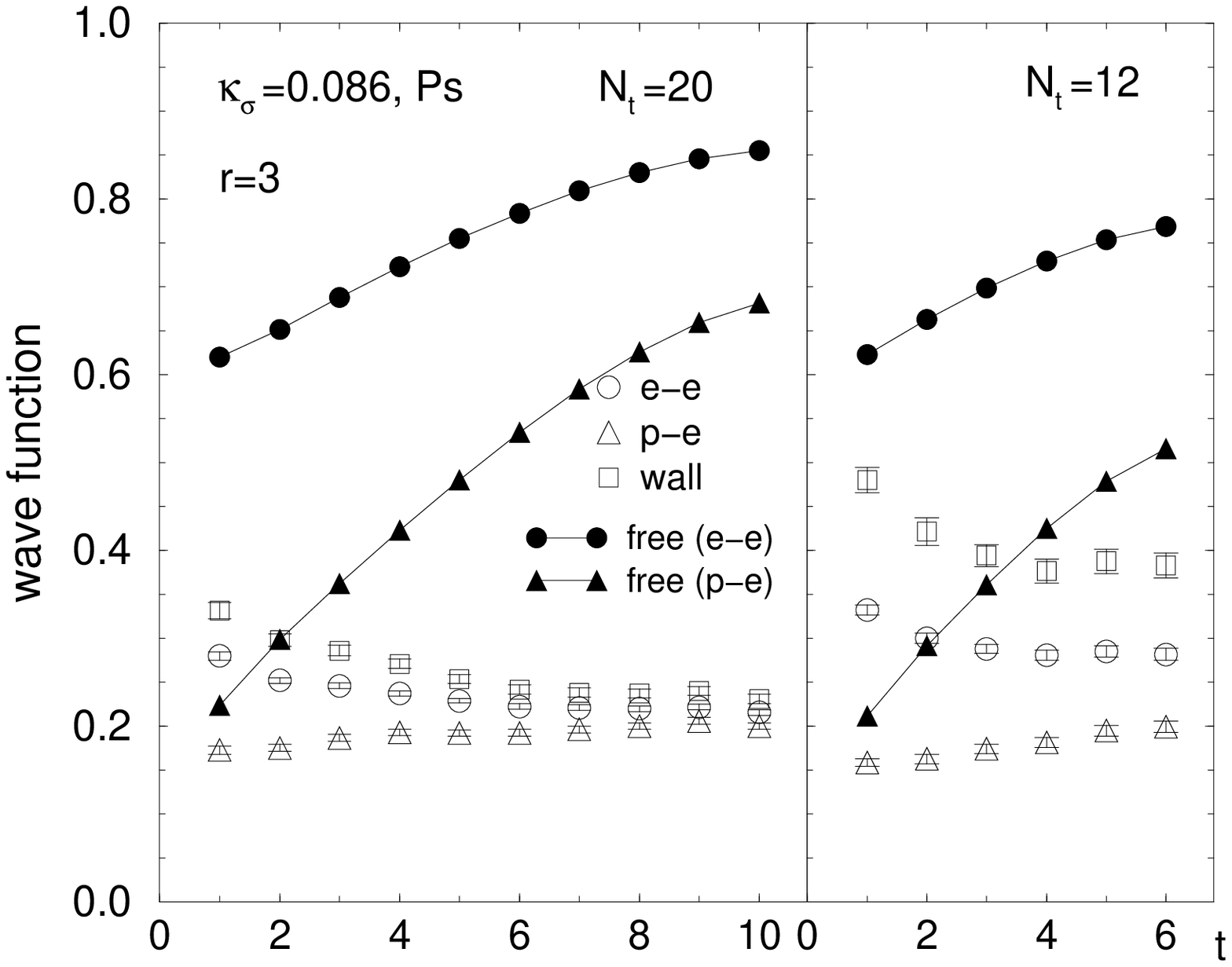,width=0.8\figwidth}
\leavevmode\psfig{file=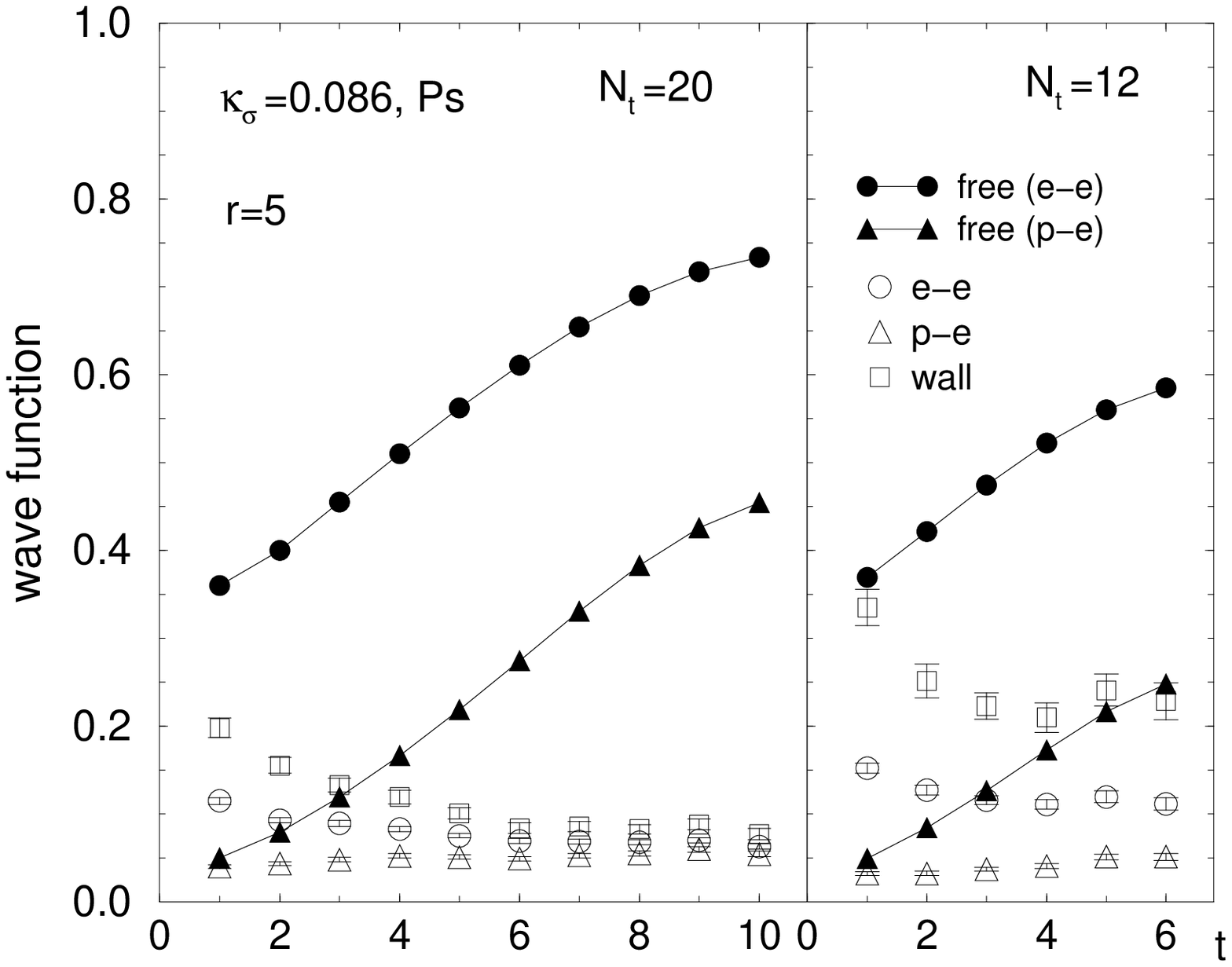,width=0.8\figwidth}
}
\caption{
Ps normalized wave function $\varphi^{(\omega)}(\vec{x},t)$
for $\vec{x}=(3,0,0)$ (left plot) and $\vec{x}=(5,0,0)$ (right plot) 
vs $t$ at $N_{\tau}=20$ and $12$ for
$\kappa_{\sigma}=0.086$ and for various sources ($point$-$point$,
$exp$-$exp$ and $wall$).
The ``free" wave function is also displayed (for $wall$ this is simply 1,
independently of $\vec{x}$ and $t$).}
\label{f.wft1}
\end{figure}

Figure~\ref{fig:wft} shows the change with $t$ of 
$\varphi^{(ee)}(\vec{x},t)$
for the Ps correlator with $exp$-$exp$ source at $N_{\tau}=20$ and $12$,
 and $\kappa_{\sigma}=0.086$. 
The ``free" meson ``wave function" $\varphi_0^{(ee)}(\vec{x},t)$
is shown for comparison.
We see now a very clear difference: The 
normalized wave function of the ``free" meson shows the expected behavior,
becoming increasingly broad  with $t$ at all temperatures. 
The measured wave function,
on the other hand, shrinks very fast from the (wider) 
distribution implied by 
the $exp$-$exp$ source at 
$t=0$ \footnote{The distance distribution of the quarks at 
$t=0$ differs somewhat
from the smearing function of the corresponding source, this does not
modify, however, the general features.} 
and stabilizes very early to a well 
defined shape with
increasing $t$. Remarkably enough, this behavior is not only seen at 
$0.93\,T_c$ ($N_{\tau}=20$) -- where, as expected, the
wave function is similar to that at $T=0$ --, but also
 at $1.5\,T_c$ ($N_{\tau}=12$): the wave function above  $T_c$ is only slightly 
wider than below.

In Figure~\ref{f.wft1} we plot
$\varphi^{(\omega)}(\vec{x},t)$ {\it vs} $t$ for
various sources, at distance
$\vec{x}=(3,0,0)$ and $\vec{x}=(5,0,0)$.
Again we show both the measured correlators and the ``free" ones.
As noticed,
a genuine wave function would be represented by ratios $\varphi(\vec{x},t)$
independent on $t$ for large enough $t$, smaller
than 1 and decreasing with $r=|\vec x|$.

From the figures we see that at 
$T=0.93 T_c$ the measured wave function approaches with increasing $t$
 a unique shape, independently of the source. 
At $T=1.5 T_c$ the $exp$-$exp$ source also appears to stabilize, while  the
$point$-$exp$ and the $wall$ sources, although  indicating some tendency
towards the same shape, do not show clear stabilization. We see therefore
here a more pronounced source dependence, very similar
to what happened with the effective masses above $T_c$.
Although the tendency to an increased 
 source dependence is a feature remembering of the ``wave function"
of the free
quark mesons, 
at all $T$ the measured wave function is very different from that given by 
free quarks. The latter, of course, have a broadening distribution
in all cases, the corresponding ratios $\varphi_0(\vec{x},t)$
increasing steadily towards 1, and show an incomparably stronger source dependence.\footnote{We use the same free quark masses (\ref{e.thm})
which were employed
for the comparison of the effective masses. 
The mass dependence is monotonous and 
heavy quarks just reproduce the initial distribution (e.g., the crosses on Fig. \ref{fig:wft} for the
$exp$-$exp$ source) at all $t$, as expected. Of course one can find some free $m_q$ and
some source to approach some of the data points, but not to reproduce  the vast 
majority of the features, in particular the shrinking
of the measured (full) wave functions with $t$ for wide initial distributions
cannot be reproduced by any free quark ansatz.}
The difference 
in behavior is particularly pregnant for the $wall$ source, where
the measured wave function shrinks strongly with $t$ while
the ``free" wave function is completely flat and independent on $t$:
$\varphi_0^{(wall)}(\vec{x},t) = 1$  for any $m_q$.  
These  features are  observed in all four measured channels.
This result is strongly indicative for the existence of (metastable) bound states 
in the mesonic channels at temperatures as high as $1.5\,T_c$, characterized 
by ``wave functions"
similar to those below $T_c$. From 
the comparison between the behavior at $0.93\,T_c$ 
and  $1.5\,T_c$ we may conclude that: a) even
at the highest $T$ the $exp$-$exp$ source
still projects on a state characterized by a strong spatial 
correlation between the quarks, quite similar to the low $T$ wave function, 
but b) with increasing $T$ above $T_c$ also other contributions in the mesonic
channels show up, without such strong spatial correlation (this
is tentatively indicated
by the increased source dependence of the wave function).  

As already remarked, 
the similarity of the stabilized $exp$-$exp$ wave functions seen at all $T$ represents 
self-consistent support for our source strategy, since the latter selects 
a given state on the basis of its spatial internal structure.

\subsection{Temperature dependence of temporal and spatial
 masses in the mesonic channels}
\label{subsec:tdep}

The discussion of the 
previous sections has shown  above $T_c$ significant temperature effects
simultaneously with the persistence of strong binding forces between
quarks. The tentative interpretation of these results is
that even above $T_c$ (metastable) bound states are
present in the mesonic channels.
In this section, assuming the existence of such states, we try
to characterize them by extracting from the propagators the temporal 
masses which would be associated with them
(would locate the corresponding peaks in the spectral functions: ``pole" masses).
They are compared with the screening masses in the same channels
and their temperature
dependence is studied.

\subsubsection*{Temporal masses}

As discussed in sections 4-5, at $N_\tau=20$ we  extract temporal masses from the correlator
$G^{(ee \rightarrow pp)}$. We use for computing the mass the last three points
near to $N_{\tau}/2$.
The fitted values are listed in Table~\ref{tab:massFT}.
Masses extracted from diagonal correlators obtained 
in the same fitting region are consistent with these values
within statistical error.

In the case of $N_{\tau}=16$ and $12$, the situation is far less clear.
Here no plateau is seen, neither for the diagonal nor for the
$ee$-$pp$ correlators.
We decide to extract masses from the latter, again by using
 the last three points near to $N_{\tau}/2$, but we should
be aware that these masses (if they do at all  characterize some states) may be
misestimated.
The resulting values are also found in Table~\ref{tab:massFT}.
They are consistent with the result of
the diagonal correlators, which suffer from large statistical
fluctuations.
We note here that  we quote statistical errors only, there are large
systematic uncertainties due, among others, to 
the smearing function dependence of the
correlators. Since we use a non-diagonal correlator (which
does not provide an upper bound for the mass) we cannot say in
which direction this uncertainty goes.

\begin{figure}[tb]
\center{
\leavevmode\psfig{file=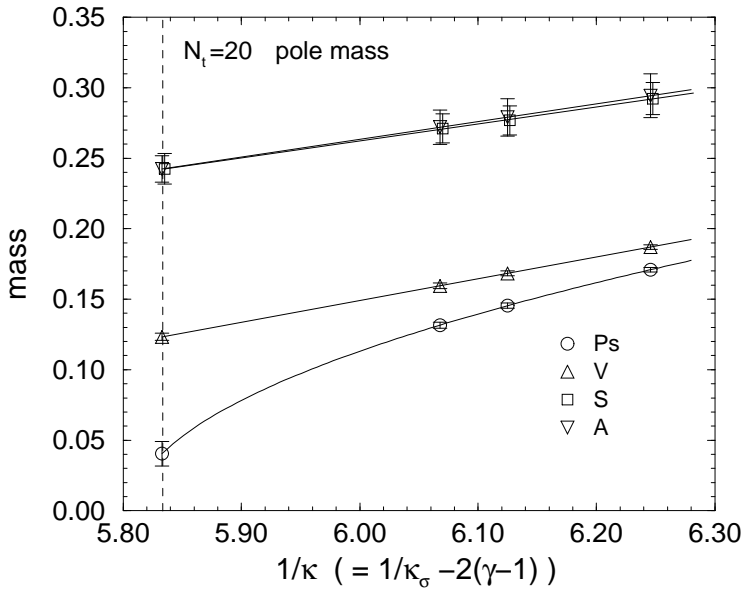,width=1.2\figwidth}}
\vspace{-0.5cm} \\
\caption{
The chiral extrapolation of the temporal masses at $N_{\tau}=20$.}
\label{fig:chext}
\end{figure}

In the next step the extracted masses are extrapolated to the chiral limit.
At $N_{\tau}=20$, in the confined phase, we extrapolate the Ps meson mass
squared linearly in $\kappa$.
For the other mesons we use linear extrapolations.
The extrapolations are shown in Fig.~\ref{fig:chext},
 they are similar to those at  $N_{\tau}=72$.
Though the Ps meson mass at the critical hopping parameter does not
completely vanish, this can be explained as an ${\cal O}(a)$ error and by the
uncertainty in the extraction of the mass.

Above $T_c$ also the Ps meson mass is extrapolated linearly. 
The  $\kappa$ dependence for all mesons is very small.
The resulting mass values at the chiral limit are listed in
Table~\ref{tab:massFT}.
Generally above $T_c$ the quark mass dependence of the meson masses is small.

\begin{table}[tb]
\begin{center}
\begin{tabular}{cccccccc}
\hline \hline
$N_{\tau}$ & $\kappa_{\sigma}$ & $m_{Ps}$ & $m_{V}$ & $m_{Ps}$ & $m_{V}$
          & $m_{Ps}^{(\sigma)}/\xi$ & $m_{V}^{(\sigma)}/\xi$ \\
\hline
20 & 0.081& 0.1708(17)& 0.1869(17)& 0.292(12)& 0.294(16)  &
                                                0.1804(18)& 0.2036(21)\\
   & 0.084& 0.1455(19)& 0.1684(18)& 0.277(10)& 0.279(13)  &
                                                0.1516(22)& 0.1816(26)\\
   & 0.086& 0.1315(22)& 0.1595(19)& 0.271(10)& 0.272(12)  &
                                                0.1354(27)& 0.1701(31)\\
   & $\kappa_c$
          & 0.040(9)  & 0.1233(25)& 0.243(11)& 0.242(9) &
                                                    -     & 0.1267(48)\\
\hline
16 & 0.081& 0.1835(14)& 0.1757(14)& 0.1877(13)& 0.1839(26) &
                                                 0.3169(34)& 0.3364(27)\\
   & 0.084& 0.1751(15)& 0.1690(14)& 0.1885(13)& 0.1731(13) &
                                                 0.3085(48)& 0.3329(31)\\
   & 0.086& 0.1723(15)& 0.1678(14)& 0.1888(13)& 0.1747(13) &
                                                 0.3036(61)& 0.3312(32)\\
   & $\kappa_c$
          & 0.1568(19)& 0.1564(16)& 0.1903(15)& 0.1804(13) &
                                                 0.2868(96)& 0.3244(41)\\
\hline
12 & 0.081& 0.2126(13)& 0.1986(14)& 0.2217(13)& 0.1981(12) &
                                                 0.4096(16)& 0.4224(21)\\
   & 0.084& 0.2100(13)& 0.1979(13)& 0.2255(13)& 0.2032(12) &
                                                 0.4036(16)& 0.4180(21)\\
   & 0.086& 0.2101(13)& 0.1996(12)& 0.2275(13)& 0.2061(12) &
                                                 0.3995(16)& 0.4148(21)\\
   & $\kappa_c$
          & 0.2062(14)& 0.2001(12)& 0.2352(13)& 0.2164(13) &
                                                 0.3868(16)& 0.4053(21)\\
\hline \hline
\end{tabular}
\end{center}
\caption{
The temporal masses and the screening masses divided by $\xi=5.3$
at finite temperature.}
\label{tab:massFT}
\end{table}

\subsubsection*{Screening masses}

Before discussing the temperature dependence of the ``masses", we briefly
describe the extraction of the spatial (``screening") masses.
Since the spatial distance is large ( $\sim$ 3 fm) we use for this
the unsmeared $pp$-$pp$ correlators and the same procedure as in the
calibration (see the Appendix).
We have verified that these masses 
are consistent with those obtained from propagators  using
 gauge invariant smearing techniques \cite{Wup}.

We measure Ps and V meson masses at all $T$. 
At zero temperature we used a.p.b.c. in all directions
and  required the masses in the space and time directions 
to represent
the same physical masses.
Thereafter we switched to periodic boundary conditions  
in the spatial directions, but this did not induce seizable changes in
the masses.

At $N_{\tau}=20$, the  spatial propagators show almost the same
behavior as at $N_{\tau}=72$.
Above $T_c$, the effective masses approach a plateau earlier than below
$T_c$,
the obtained values are therefore more reliable than in the confining phase.

Masses are again extrapolated to the chiral limit.
Also here, as for temporal masses, due to uncertainties and
partly to ${\cal O}(a)$ effects
the naive extrapolation of the Ps meson mass squared does
not vanish exactly at  $1/\kappa_c$.

\begin{figure}[tb]
\center{
\leavevmode\psfig{file=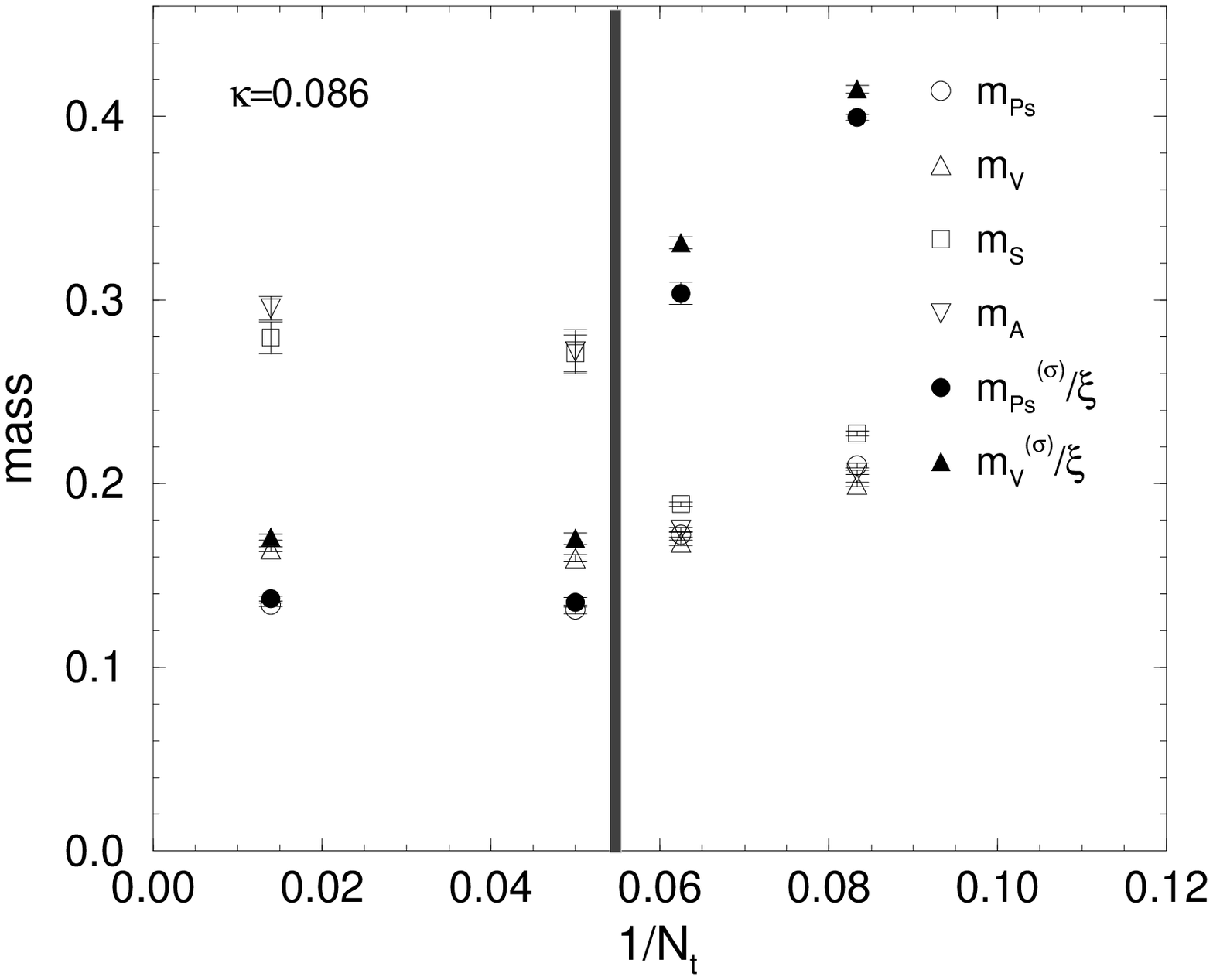,width=\figwidth}}
\center{
\leavevmode\psfig{file=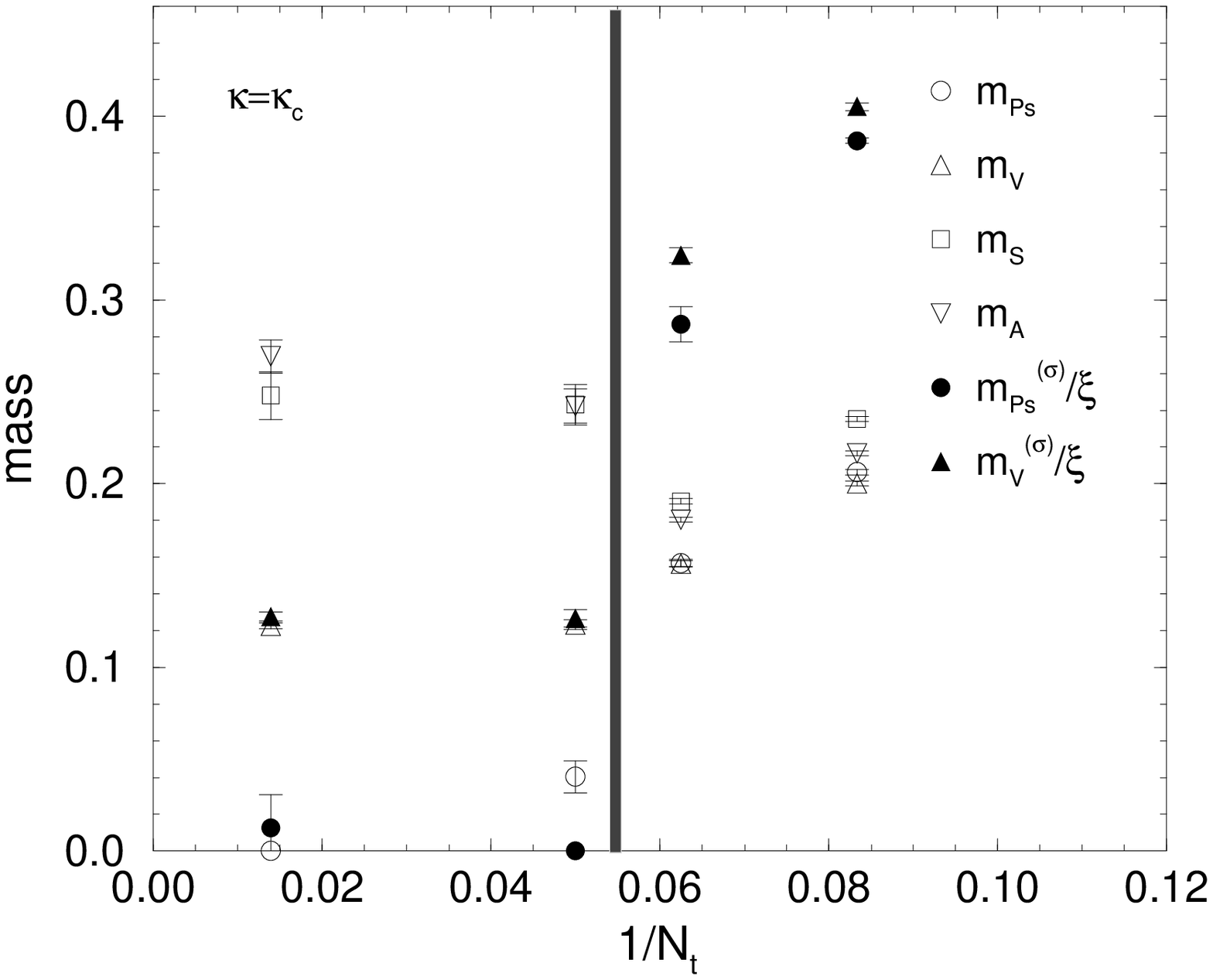,width=\figwidth}}
\caption{
Temperature dependence of masses (in $a_{\tau}^{-1}$)
at $\kappa_{\sigma}=0.086$ (top) and in the chiral limit (bottom).
Full (open) symbols correspond to spatial (temporal) masses.
The grey vertical line roughly represents $T_c$.
}
\label{fig:Tdep}
\end{figure}

\begin{figure}[tb]
\center{
\leavevmode\psfig{file=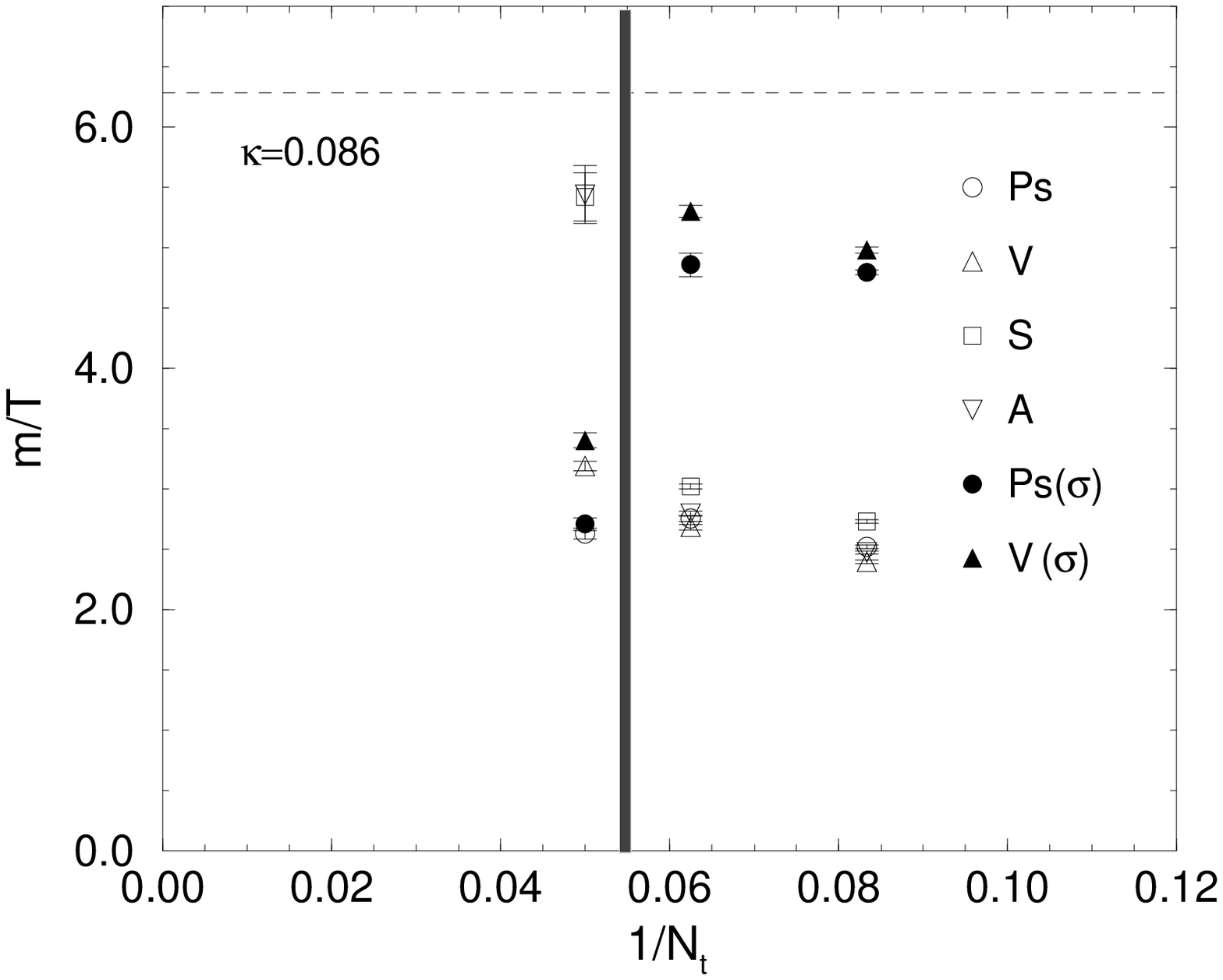,width=\figwidth}}
\center{
\leavevmode\psfig{file=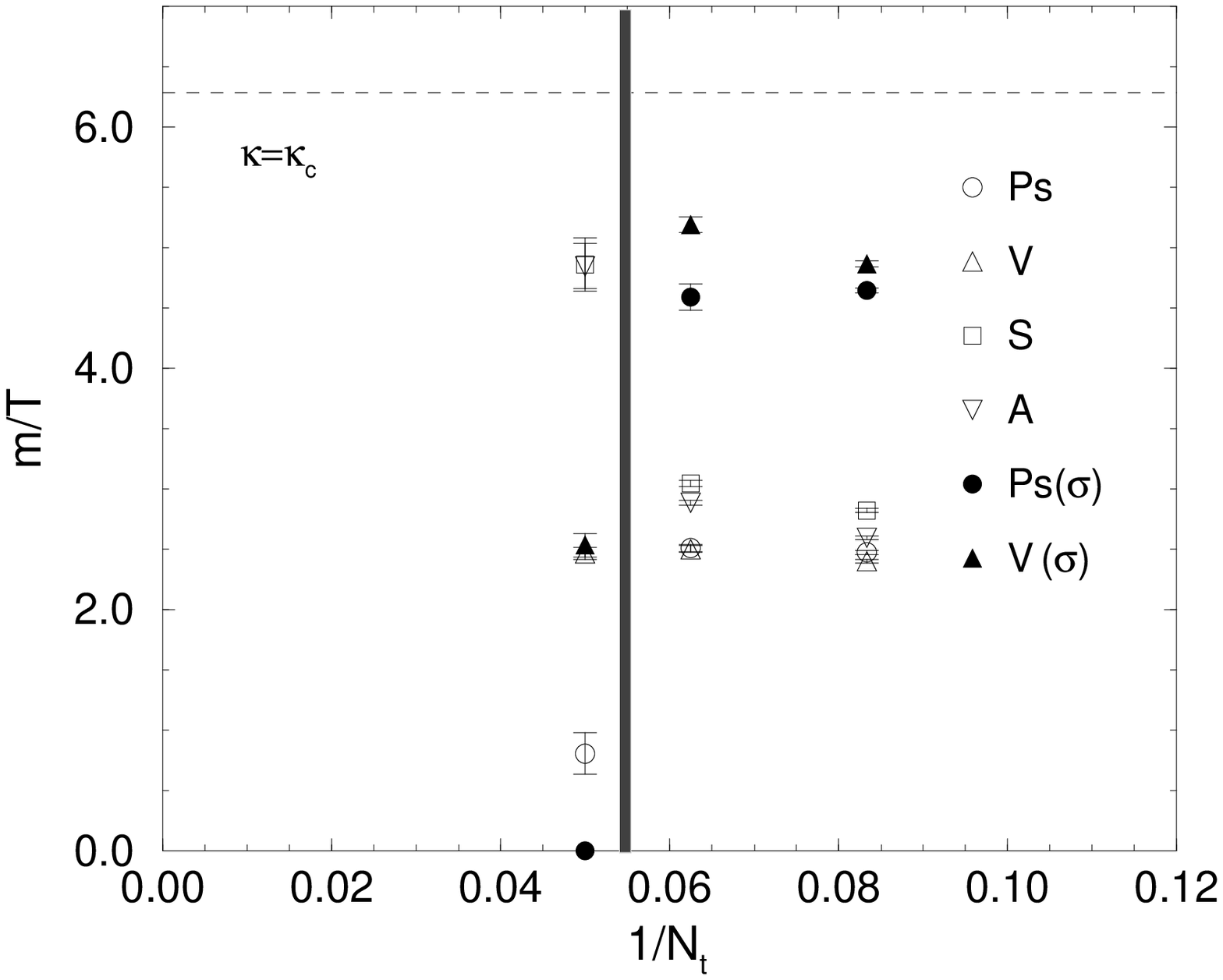,width=\figwidth}}
\caption{
$m/T$ at $T \simeq 0.93\,T_c,\,1.15\,T_c$ and $1.5\,T_c$ for
$\kappa_{\sigma}=0.086$ (top) and in the chiral limit (bottom).
Full (open) symbols correspond to spatial (temporal) masses.
The grey vertical line roughly represents $T_c$.
The dashed horizontal line corresponds to twice the lowest Matsubara
frequency. }
\label{f.mot}
\end{figure}

\begin{figure}[tb]
\center{
\leavevmode\psfig{file=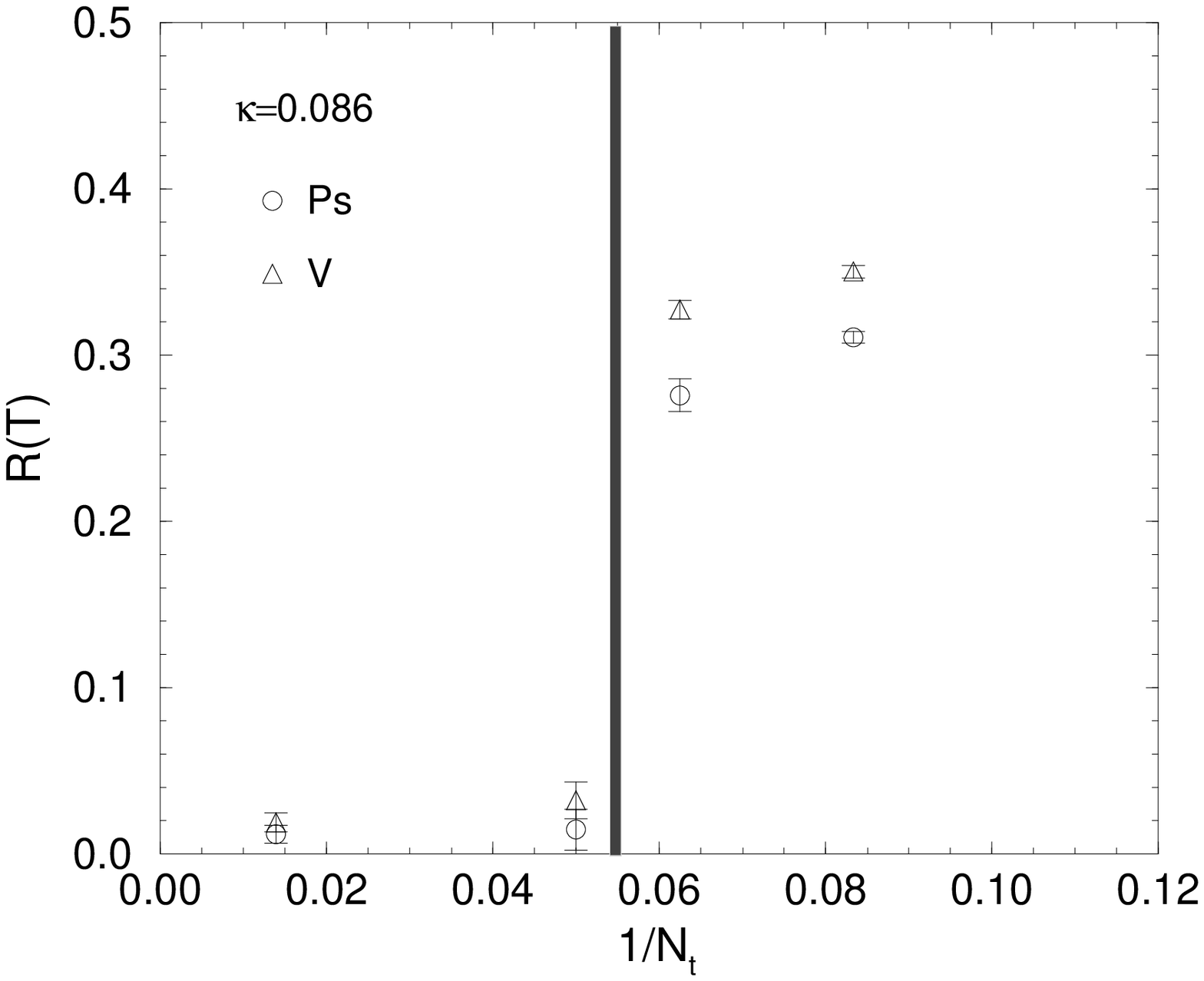,width=\figwidth}}
\caption{The phenomenological parameter $R$ eq. (\ref{e.R}) as
function of $T$. 
The grey vertical line roughly represents $T_c$.
}
\label{fig:R}
\end{figure}

\subsubsection*{Temperature dependence}

The temperature dependence of temporal and screening masses
is summarized in Fig.~\ref{fig:Tdep}.
The values are given in units of $a_{\tau}^{-1}$.
The horizontal axis is $1/N_{\tau}$ and we have 
$T=1/N_{\tau} a_{\tau}$ with $a_{\tau}^{-1}=4.5(2)$ GeV. The four points
($N_{\tau} = 72,\,20,\,16$ and $12$) correspond to the temperatures
$T \simeq 0,\,0.93\,T_c,\,1.15\,T_c$ and $1.5\,T_c$.
The vertical gray line roughly corresponds to the critical temperature.

In the confining phase, the temporal and  screening (spatial) masses
coincide, however above $T_c$ they become increasingly different.
This is to be expected, since the former (whether they represent
bound states or not) are related to propagation in plasma with the 
transfer matrix of the original problem, 
while the latter correspond to 
a $T=0$ problem with {\it asymmetric} finite size effects,
strongly increasing with the temperature (one of the ``spatial" direction
becomes squeezed as $1/T$).
In agreement with other works \cite{scr},
screening masses above $T_c$ are $\propto T$ 
and close to  twice the lowest Matsubara
frequency (the lowest quark momentum in the squeezed direction with a.p.b.c.), 
although remaining below it for all $T$. On Fig.~\ref{f.mot} we
plot the ratios $m/T$ for the three non-zero temperatures together with twice the
lowest Matsubara frequency for comparison. The temporal masses above $T_c$
are also proportional with $T$ but with a 
significantly smaller slope. The slight decrease of the Ps and  V  temporal masses ($\times 1/T$)
above $T_c$ in the upper plot of Fig.~\ref{f.mot} is due to 
the large quark mass in the simulation,
 which produces a term $\propto 1/T$ in this plot. 
This decrease disappears in the chiral limit (lower plot).

As a phenomenological parameter to succinctly quantify this
behavior we introduce:
\be
R = \frac{ m^{(\sigma)} - m^{(\tau)} }{ m^{(\sigma)} + m^{(\tau)} }\
\towh{}{\longrightarrow}{\sss \rm P.Th.}
\ 1 - \frac{2m_q }{\pi T}+...\,,\ \ \
m_q  \ll T\ll a_{\tau}^{-1} .
\label{e.R}
\ee
\no Since  at high $T$ the quarks are expected
to exhibit an effective (temporal)
mass $m_q^{eff} \sim g^2T/\sqrt{6}$ \cite{fea1}, 
$R \sim 1 - 0.26\,g^2$ for $T \gg T_c$.
From  Fig.~\ref{fig:R} we see that our data tend toward this 
regime but are still well below it even at $T=1.5\,T_c$. 
Notice that because of large lattice artifacts 
in our data mass ratios are more reliable than absolute values.

Let us note again that above $T_c$ all four channels show almost the same masses.
In the present quenched calculation, the chiral symmetry is not involved
in the dynamics and the phase transition is the deconfining transition.
Nevertheless, the chiral symmetry seems to be restored, which indicates a
close relation between the two viewpoints of the QCD phase transition.
This agrees with the old observation that the chiral condensate also feels
the deconfining transition of pure gauge theory -- see, e.g., \cite{kog}.

\section{Conclusions and outlook}

In this
 quenched QCD analysis the changes of the meson properties with temperature
appear to be small below $T_c$.
Above $T_c$ we observe apparently
opposing features:
On the one hand, the behavior of the $t$-propagators, in particular
the change in the
ordering of the mass splittings could be accounted for by contributions from
free quark propagation  in the mesonic channels,
which would  also explain
the variation of $m^{(\tau)}_{eff}(t)$ both with $t$ and with the source.
 On the other hand, the behavior of the wave functions obtained from the 
  4-point correlators
suggests that there can be
 low energy excitations in the mesonic channels   above $T_c$
 appearing as metastable bound states which replace
 the low temperature mesons. 
They would be characterized by a mass 
giving the location of the corresponding bump in the
spectral function. 
In this case the variation of $m^{(\tau)}_{eff}(t)$ with $t$ and with the source
would indicate a resonance width for these states, increasing with $T$.
Our treatment of the
low energy states using the $exp$-$exp$ source
introduces, however,  uncertainties which do not
allow for quantitative conclusions.
As we have seen this source is indeed
slightly too wide and 
includes some contributions from excited states compensating each other 
in the $ee$-$pp$ correlator  at $T < T_c$.
At high $T$ 
the ``effective" mass becomes therefore increasingly ambiguous. 
Remember, however, that our source is not
chosen arbitrarily but selects a state according to the internal structure of 
the latter on the basis of the similarity with the wave function of the
$T=0$ mesonic ground state.
What we find is that at all temperatures there is a tendency for stable spatial
correlation between quarks with a shape similar to the $T=0$ wave function.
It seems therefore reasonable to hypothesize that 
at all temperatures this source finds a low energy mode characterized by
strong, stable spatial correlations between the quarks, and that the properties of
the propagators taken with this source will reflect within some uncertainties the
properties of this mode.

We see clear signal for chiral symmetry restoration starting early
above $T_c$. Since this is
a quenched calculation, this effect is completely due to the gluonic dynamics.
Above $T_c$ the screening masses
 increase faster than the (temporal) masses, remaining however 
 clearly below the free gas limit.
The exact amount of splitting among the channels and the
precise ratio between $m^{(\tau)}$ and $m^{(\sigma)}$ may, however,
 be affected also by uncertainties in our $\xi$ calibration,
 in the definition of the source etc:  
 Especially the temporal masses might be misestimated (if they are
 at all well defined). But the semi-quantitative picture of much 
(and, with $T$, increasingly) steeper ``screening" propagators (as
compared with temporal ones) is undoubtful. 

A possible physical picture is this: 
Mesonic excitations are present above $T_c$
(up to at least $1.5 T_c$) as unstable modes (resonances), in interaction
with unbound quarks and gluons. They may be realized as collective states,  
by the interaction of the original mesons with new effective degrees of 
freedom in the thermal bath, or as metastable bound states (of thermal quarks?).
 To the extent that our results can be
interpreted as supporting this picture, we should note that:\par
-\ the ``temporal masses" 
of these putative states in the pseudo-scalar and vector channels connect 
smoothly  to the ``pole" 
masses of the
mesons below $T_c$ (the
S and A correlators change more significantly, in accordance with
the chiral symmetry restoration),\par
-\ the increase of the temporal meson masses with $T$ is to a certain
extent similar to that of the quark 
 thermal mass if we assume $g^2 \sim 1$ (while the spatial masses
increase much faster),\par
-\ the wave functions characterizing these states are very similar to those of the 
genuine mesons below $T_c$,\par
-\ since there is no pair creation (no dynamical quarks) a ``decay channel" for 
these putative resonances may be $meson \rightarrow q_{th}\ +\ {\bar q}_{th}$
where by $q_{th}\,+\,{\bar q}_{th}$ we indicate some strongly interacting 
but unbound 
quark states,\par
-\ the incomplete (quenched) dynamics of this simulation seems to already 
provide enough
interaction to account (besides for mesonic states below $T_c$) for chiral symmetry
restoration and binding forces above $T_c$; nevertheless effects of dynamical quarks
of small enough masses may add further features to this picture. 

Although our results
are consistent with the above picture, there may be also other possibilities
(cf \cite{dTar,biel1}, cf \cite{hild} and references therein).
We also see agreement with  the earlier study of meson propagators including 
 dynamical quarks
(but  without  wave function information) \cite{biel1},
which finds masses and (spatial) screening masses $\propto T$
above $T_c$ 
and indication for QGP with ``deconfined, but strongly
interacting quarks and gluons".
 Investigations 
of mesonic correlation functions in HTL approximation \cite{fri00}
show that the observed effects cannot be reproduced  from perturbative
thermal physics. 
 The wave function analysis 
in our work indicates 
in fact 
that the strong interactions between the thermal gluons and quarks may
even provide 
binding forces which partially correlate the latter in space.

The complex, 
non-perturbative structure of QGP (already indicated by 
 equation of state studies up to far above $T_c$ \cite{LattFT},
see also \cite{Boy96}) 
is thus also confirmed by our analysis
of general mesonic correlators.
From our more extended study however, especially from the, here for the first time
investigated, spatial correlations between quarks propagating in the temporal 
direction at $T > T_c$ (wave functions),
the detailed low
energy structure of the mesonic channels appears to present
further interesting, yet unsolved aspects and therefore provide an
exciting and far from
trivial picture of QGP in the region up to (at least) $1.5T_c$. 
Further work is needed to remove
the uncertainties still affecting our analysis. This concerns
particularly the $\xi$ calibration and the
question of the definition of hadron operators at high $T$,
which appear to have been the major deficiencies, besides the smaller lattices,
affecting earlier results \cite{HNS93a}.
We are also trying to extract
information directly about the
spectral functions \cite{hsp}. Finally we are aiming at performing a
full QCD analysis in the near future.
\par\bigskip

\no {\bf Acknowledgments:} We thank
JSPS, DFG and the European Network ``Finite Temperature Phase
Transitions in Particle Physics" for support. H.M. thanks T. Kunihiro 
and H. Suganuma  and I.O.S. thanks F. Karsch and J. Stachel
for interesting discussions. 
H.M. also thank the Japan Society for the Promotion of Science
 for Young Scientists for financial support.
O.M. and A.N. were supported by  the Grant-in-Aide for Scientific
Research by the Ministry of Education and Culture, Japan
(No.80029511) and  A.N. was also  supported by the Grant-in-Aide 
(No.10640272). M.G.P. was supported by CICYT under grant AEN97-1678.
The calculations have been done on the AP1000 at Fujitsu Parallel
Comp. Res. Facilities and the Intel Paragon at the
Institute for Nonlinear Science and Applied Mathematics, Hiroshima University.

\section*{Appendix: Calibration of anisotropy parameters}

To determine the gauge field anisotropy $\xi$, we use the ratios of the
spatial-spatial and spatial-temporal Wilson loops
\cite{TARO97,EKS98,Kla98}:
\begin{eqnarray}
 R_{\sigma}(r,x) &=& W_{\sigma \sigma} (r+1,x) / W_{\sigma \sigma} (r,x),
 \nonumber \\
 R_{\tau}(r,t)   &=& W_{\sigma \tau} (r+1,t) / W_{\sigma \tau} (r,t).
\end{eqnarray}
Then the matching condition (\ref{e.phiso}) is
\begin{equation}
 R_{\sigma}(r,x) = R_{\tau}(r,t=\xi\, x).
\end{equation}
80 configurations are used for this analysis.

In the determination of $\xi$, we vary the minimum value of $r\times x$
(with corresponding choice of $t$), where $r\leq5$
($r=1$ and $x=1$ are  not used to avoid  short
distance effects).
The largest value of $x$ for each $r$ is chosen with consideration to
the statistical errors.
We obtain
$\xi = 5.397 (22)$ ($r \times x \geq 4$),
$5.340 (40)$ ($r \times x \geq 6$),
$5.248 (101)$ ($r \times x \geq 8$) and take therefore
 $\xi = 5.3(1)$
in the following.

To determine the lattice spacings, the heavy quark potential is measured.
The extracted value of the string tension $\sqrt{\sigma}=0.480(23)$ together with
physical value $\sqrt{\sigma_{phys}} = 427$ MeV gives
the cutoffs $a_{\sigma}^{-1} = 0.85(3)$ GeV
and $a_{\tau}^{-1} = 4.5(2)$ GeV.
The spatial extent of the lattice of about 3 fm ($\sim 4$ times $1/T_c$)
is considered sufficiently
large to treat hadronic correlators.

We then proceed to the fermionic calibration.
We fix the value of $\kappa_{\sigma}$ and vary $\gamma_F$ to find out
the value which gives the same anisotropy $\xi$ as for the gauge field.
We define the fermionic anisotropy using
 correlators in temporal and spatial directions, expected to behave
at large distances like
\be
 G_{\tau,M}^{(pp  \rightarrow pp)}(t) =
                   \langle \Phi_M^{(pp)}(t)\,
\Phi_M^{(pp)}(0)^{\dag} \rangle
  \towh{}{\longrightarrow}{\sss t \rightarrow \infty}
  C_{\gamma}^{(\tau)}\, \exp(-m_{M}^{(\tau)} t)
\ee
and the same expression with time replaced by one of the spatial directions $z$,
behaving as $C_{\gamma}^{(\sigma)}\, \exp(-m_{M}^{(\sigma)} z)$
for large $z$.

\begin{figure}[tb]
\center{
\leavevmode\psfig{file=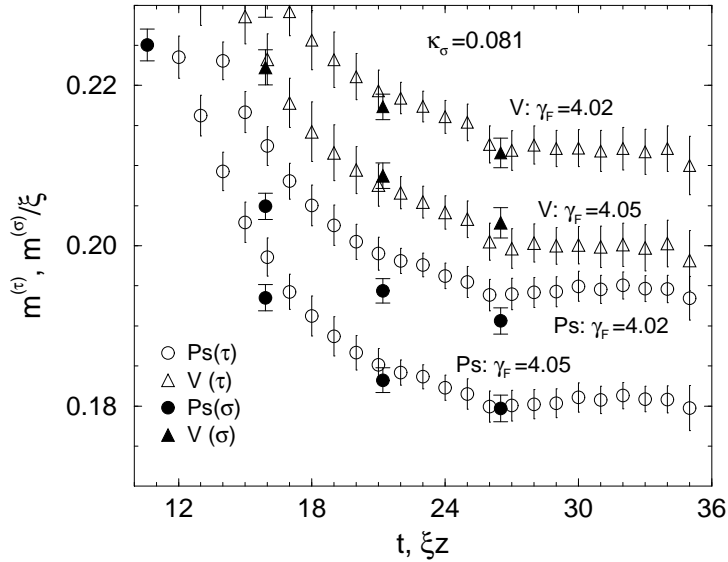,width=\figwidth}}
\caption{
The effective masses of correlators in temporal
and spatial directions for $\kappa_{\sigma}=0.081$
with two values of $\gamma_F$, 4.02 and 4.05.}
\label{fig:qcalib1}
\end{figure}

In the calibration, we measure the pseudo-scalar ($\gamma_{\rm M}=\gamma_5$)
and the vector ($\gamma_1$, $\gamma_2$) meson correlators.
Here we adopt  anti-periodic boundary conditions
in all four directions, hence
 the  expected behavior is a hyperbolic cosine. The physical
isotropy condition (\ref{e.phiso}) is then applied to the
effective masses.
Figure \ref{fig:qcalib1} shows  $m^{(\tau)}_{eff}(t)$
obtained by solving (\ref{e.efem}) and the corresponding
 $m^{(\sigma)}_{eff}(z)$ for $\kappa_s=0.081$ and
 with two values of $\gamma_F$. On the figure
$m^{(\sigma)}_{eff}(z)$ is divided by $\xi=5.3$ to be compared
with $m^{(\tau)}_{eff}(t)$ (i.e., it is given in units $a_{\tau}^{-1}$).

For $\gamma_F=4.05$, the spatial effective mass divided by $\xi=5.3$
coincides with the temporal one. Although the former shows no plateau because of
the small
number of spatial sites,
the temporal effective mass, which is finer spaced,
does reach a plateau in the large $t$ region.
It is consistent to expect that if both masses agree (after $\xi-$rescaling)
in the region where the temporal mass shows a plateau, the spatial mass
is also dominated by the ground state.
We therefore determine $\xi_F$, the fermionic anisotropy, as the ratio
of the spatial effective mass at $z=5$ and the fitted value
of the temporal mass in the interval $t=27$-$36$.
The value of $\xi_F$ and the extracted masses are confirmed by studying correlators with smeared
operators which do reach plateaus much earlier.
The smearing procedure of the correlators in the temporal direction is
described in detail in the next section.
For the correlators in spatial directions, we apply the gauge invariant
smearing technique \cite{Wup}
(since the configurations are fixed to the Coulomb gauge).

Though these calculations are carried out with periodic boundary conditions
in spatial directions, the dependence of masses on the kind of boundary
conditions is sufficiently small on the present lattice.

\begin{figure}[tb]
\center{
\leavevmode\psfig{file=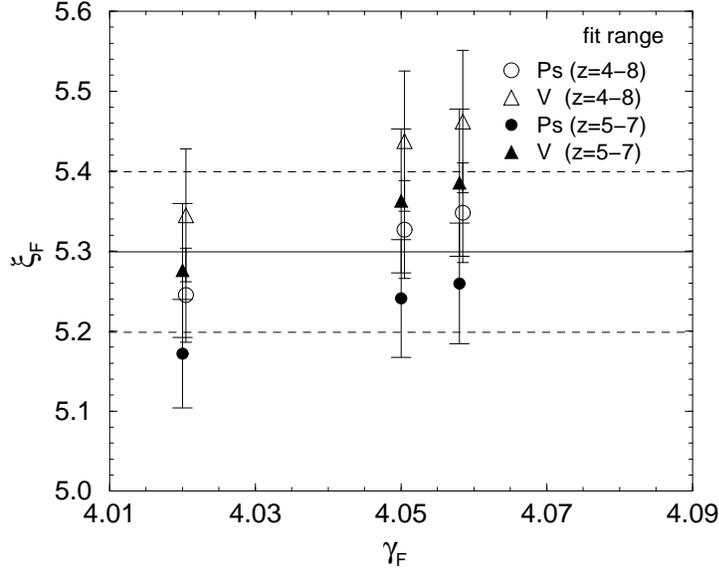,width=\figwidth}}
\caption{
$\xi_F$'s determined from the Ps and V correlators
for $\kappa_{\sigma}=0.081$ with various values of $\gamma_F$.
}
\label{fig:qcalib2}
\end{figure}

Figure~\ref{fig:qcalib2} shows the dependence of $\xi_F$ on $\gamma_F$.
The values of $\xi_F$ from Ps and V mesons are slightly different, but
consistent within the present accuracy.
We adopt the averaged value of Ps and V channels
and estimate the error as their difference.

We use three sets of ($\kappa_{\sigma}$, $\gamma_F$):
(0.081, 4.05), (0.084, 3.89) and (0.086, 3.78).
In Table~\ref{tab:parameters}, these values are listed together with
the number of configurations used for calibration.
For the second set, two values of $\gamma_F$ are tried.
The meson masses quoted in the table are determined in section 4
using smeared correlators.

Another procedure to calibrate the fermionic action
using the dispersion relation is proposed in \cite{Kla99}.
In the present case, however, the  procedure used above
seems more appropriate, since comparison of pole and screening masses at finite
temperature is one of the important goals of this work.

\end{document}